\documentclass[letterpaper]{aa}  
\usepackage{graphicx}
\usepackage{txfonts}
\usepackage{lscape}
\usepackage{epsfig}
\usepackage{times}
\usepackage{graphicx}
\usepackage{natbib}
\bibpunct{(}{)}{,}{a}{}{}
\begin{document}
\title{The near--infrared luminosity function\\
 of cluster galaxies beyond redshift one\thanks{Based in part on observations obtained at the European Southern 
Observatory using the ESO Very Large Telescope on Cerro Paranal (ESO program 166.A-0701). 
Based in part on observations obtained at the Hale Telescope, Palomar Observatory,
as part of a continuing collaboration between the California Institute
of Technology, NASA/JPL, and Cornell University.} }

   \author{V. Strazzullo\inst{1,2} \and P. Rosati\inst{2}
\and S. A. Stanford\inst{3,4} \and{C. Lidman}\inst{5}
\and M. Nonino\inst{6} \and \\R. Demarco \inst{7}
 \and P. E. Eisenhardt\inst{8}\and S. Ettori\inst{9}
\and V. Mainieri\inst{10} \and S. Toft\inst{11} 
          }

   \offprints{V. Strazzullo, vstrazzu@eso.org, strazzul@na.astro.it}

   \institute{Dipartimento di Scienze Fisiche, Universit\`a degli Studi di Napoli ``Federico II'', via Cinthia, I--80126 Napoli, Italy
\and
European Southern Observatory, Karl--Scwarzchild--Strasse 2, D--85748 Garching, Germany
\and
Department of Physics, University of California at Davis, 1 Shields Avenue, Davis, CA 95616, USA
\and
Institute of Geophysics and Planetary Physics, LLNL, Livermore, CA 94551,  USA
\and
European Southern Observatory, Alonso de Cordova 3107, Casilla 19001, Santiago, Chile
\and
Instituto Nazionale di Astrofisica, Osservatorio Astronomico di Trieste, via G.B. Tiepolo 11, I--34131, Trieste, Italy
\and
Department of Physics and Astronomy, Johns Hopkins University, 3400 N. Charles Str., Baltimore, MD 21218, USA 
\and
Jet Propulsion Laboratory, California Institute of Technology, MC 169--327, 4800 Oak Grove Drive, Pasadena, CA 91109
\and
Istituto Nazionale di Astrofisica, Osservatorio Astronomico di Bologna, via Ranzani 1, I--40127 Bologna, Italy
\and
Max--Planck--Institut f\"{u}r extraterrestrische Physik, Postfach 1319, D--85748 Garching, Germany
\and 
Department of Astronomy, Yale University, P.O. Box 208101, New Haven, CT 06520--8101, USA
             }


 
  \abstract
   { }
   {We determined the $K_s$ band luminosity function (LF), and inferred the corresponding stellar mass function, of cluster galaxies at redshift $z\simeq 1.2$,  using near--infrared images of three X--ray luminous clusters at $z=1.11,1.24,1.27$.}
   {The composite LF was derived down to M$^{*}$+4, by means of statistical background subtraction,  and is well described by a Schechter function with $K_s^{*}=20.5^{+0.4}_{-1}$ and $\alpha=-1.0^{+0.2}_{-0.3}$. Using available X--ray mass profiles we determined the M/L ratios of these three clusters, which tend to be lower than those measured in the local universe.
Finally, from the $K_s$ band composite LF we derived the stellar mass function of cluster galaxies.}
   {With these data, no significant difference can be seen between the cluster galaxies LF and the LF of field galaxies at similar redshift.
We also found no significant evolution out to $z \simeq 1.2$ in the bright ($<$ M$^{*}$+4) part of the LF probed in this study, apart from a brightening of  $\simeq 1.3$ mag of the characteristic magnitude of the high redshift LF. We confirm, and extend to higher redshift, the result from previous work that the redshift evolution of the characteristic magnitude M$^{*}$ is consistent with passive evolution of a stellar population formed at $z>2$.
}
   {The results obtained in this work support and extend previous findings that most of the stars in bright galaxies were formed at high redshift, and that $K_s$--bright ($M>10^{11} M_{\odot}$) galaxies were already in place at $z\simeq 1.2$, at least in the central regions of X--ray luminous clusters. Together with recent results on the field galaxies stellar mass function,  this implies that most of the stellar mass is already assembled in massive galaxies by z $\simeq 1$, both in low and high density environments.}


   \maketitle
%

\section{Introduction}

Galaxy clusters are rare systems forming in the highest density peaks of 
large scale structure. In these special regions galaxy formation
and evolutionary processes are expected to be faster with respect to the low
density fields, thus making galaxy clusters a biased environment. On
the other hand, clusters of galaxies, particularly at high redshift
($z \simeq 1$), provide a very convenient place for studying the
evolution of massive galaxies.  Not only do they contain high numbers of
such objects, but these objects turn out to be so evolved (already at
$z \simeq 1$) that they show a colour--magnitude sequence as clear as at
lower redshifts.  Thus evolved galaxies in distant clusters
can be easily identified even
without complete spectroscopic follow--up.

The study of massive galaxies has a relevant role in constraining
galaxy formation and evolution models, as different models provide
different predictions for their assembly (in particular in the
redshift range [$0 \div 1$]).  They could have rapidly formed their
stars at high redshift and at the same time assembled their stellar
mass, and then simply evolved passively as their stars aged.
Alternatively, massive galaxies could have assembled on a longer time
scale in a process of continuous merging of smaller units until
redshift $< 1$.  Comparison of these different scenarios has proven to
be a difficult task: even if merging galaxies are observed, the
relevance of the merging process in galaxy evolution and especially
the epoch at which major mergers occur is still debated.

Colours and spectra of massive galaxies at $z \simeq 1$ show that
there is a significant population of such systems already hosting
mainly evolved stellar populations, both in the field (see below) and
in clusters (\citet{stanford1997, stanford1998, vandokkum1998,
benitez1999, depropris1999, rosati1999, stanford2002, blakeslee2003,
vandokkum2003, kodama2004, lidman2004, delucia2004, holden2005b,
holden2005c, tanaka2005}). In such studies, the presence of evolved
stellar populations is generally inferred from fundamental plane or
colour--magnitude sequence evolution. These studies indicate that most
of the stars in massive galaxies were formed at $z >2$. At the same time
several of them point out that the less massive the galaxy is,  the more
likely is the presence of a younger component in its stellar
population -- the so called ``downsizing'' \citep{cowie1996} in galaxies hosting star formation.

It is not possible, however, on the basis of spectrophotometric
analysis only, to rule out the possibility that these galaxies formed
via merging of smaller galaxies with already evolved stellar
populations even a short time before being observed.  For instance, a
passively evolving zero--point of the colour--magnitude relation does
not imply that the galaxies formed long ago, but that the stars in the
galaxies formed at high redshift, possibly in smaller progenitors.  

In other words, while the underlying stellar populations can place
constraints on the details of the star-formation history, they cannot
tell when a galaxy assembled.  In fact, even if these massive galaxies
appear to be passively evolving, several studies have noted that to
some extent they can still be forming (or recently have formed) stars
(for instance
\citet{nakata2001,vandokkum2003,holden2005,demarco2005,jorgensen2005}),
likely implying merging (as observed for instance by
\citet{vandokkum1999} and \citet{tran2005}) or anyway subsequent
episodes of star formation. Such secondary episodes of star formation
are probably correlated to cluster--related processes (accretion of
field galaxies or groups and cluster merging), since galaxies which
exhibit these features are often located outside of the cluster core,
or in regions of lower X--ray luminosity.  Moreover, the redshift
evolution of the brightest cluster galaxies (BCGs) is peculiar and
exhibits a large scatter at increasingly high redshift, so that at
least in some cases merging could be required to make them evolve into
local BCGs (see for instance \citet{ellisejones}). Even if it is
difficult to trace a common ``BCG evolutionary path'', due to the
intrinsically peculiar nature of these galaxies, some high redshift
clusters (including Cl1252 \citep{rosati2004b,blakeslee2003} and
Cl0848 \citep{vandokkum2001}) show signs of interactions or clear
ongoing merging between few massive galaxies which could lead to the
formation of a cD.

However, even if the bulk of the stars
had similar ages in the two formation scenarios (i.e. star formation
 occurring at the same early epoch in both), the epoch of assembly of
the final mass observed locally in massive galaxies is different in
the two cases.  If merging is a relevant process in the look--back time
range that we can probe with our observations, looking at
progressively higher redshift one should see the number of massive
objects decreasing as they break up into their progenitors; this would
cause an evolution in the shape of the mass function of galaxies.

The differences in the predictions of the two formation scenarios have
recently become less extreme, partly due to the higher redshift peak
of the merging activity in $\Lambda$CDM models as compared to standard
CDM initially considered, but also to different ad--hoc recipes for
the star formation adopted in the hierarchical merging models.

It has recently become more evident that, both in clusters (references
mentioned above) and in the field
(\citet{glazebrook2004,mccarthy2004,fontana2004,saracco2004,
cimatti2002k20, franx2003}), a significant population of massive
galaxies is already in place at $z \simeq 1 \div 2$. However, see also
\citet{vandokkum2005}, finding that a considerable fraction of 
a nearby bulge--dominated galaxy sample, recently experienced 
a merging episode involving more than 20\% of its final mass.
The stellar mass function of bright galaxies shows only a mild
evolution up to redshift 1 , close to the prediction of simple pure
luminosity evolution \citep[e.g.][]{fontana2004}.

The comparison with recent semi--analytical models however shows that
different renditions predict very different evolution, especially at
higher redshift (i.e. results are very sensitive to the chosen model
ingredients), and most of them under--produce very massive galaxies
(more severely the higher the redshift) even when reproducing the
stellar mass function around M$_{*}$ -- however, see also recent
results from \citet{bower2005}. At the same time, \citet{nagamine2005}
show that with recent hydrodynamical simulations 
they can account for $\simeq 70\%$ of the total stellar mass at
$z =0$ already being formed by $z=1$.

Since a direct measure of the mass function is too difficult at high
redshift for a reasonably large galaxy sample including faint objects,
the near--infrared galaxy luminosity function (LF) represents an useful
cheaper surrogate. The galaxy LF is a first order description of a
galaxy population (density of galaxies as a function of their
luminosity). Despite (or because of) its conceptual simplicity the LF
has been for many years one of the most popular tools for the
interpretation of galaxy observations at all redshifts and in very
different environments. The comparison of the LF at different
redshifts constrains models of galaxy formation and evolution
\citep{kauffmannecharlot1998b}, while the comparison of the LF in low
and high density environments probes the relevance of the
environmental effects on the galaxy populations.  The LF is 
historically best studied in rich clusters of galaxies, which provide
large numbers of galaxies at the same distance and, at low redshift,
with high contrast against the background, allowing an efficient
identification of cluster members with small contamination from
background galaxies.  At higher redshift, the faint luminosities and
the substantial background contamination makes the LF  determination more
uncertain. However, the steadily increasing data quality, and the quest
for strict constraints on galaxy evolutionary models, have made the study of
the LFs in high redshift clusters and fields a popular topic.

In this work, we determine the LF of distant ($z >1$) cluster galaxies
in the near--infrared (NIR).  NIR galaxy samples are particularly well suited for studying galaxy evolution.  Apart from
advantages such as the smaller effect of dust extinction
(as compared to bluer wavelengths), and the k--corrections relatively
insensitive to galaxy type, they provide a relatively good estimate of the
stellar mass in galaxies up to redshift $z \sim 2$.  Therefore, 
near--infrared luminosity functions can trace the stellar mass
function more effectively than bluer band LFs,  which are more
sensitive to the star formation histories of the galaxies.

While LFs for cluster galaxies at low redshift ($z \leq 0.2 \div 0.3$) have 
been determined for a large number of clusters, allowing detailed
discussion of the features and the separate contributions of
different galaxy populations down to very faint magnitudes, the
determination of the LF with comparable accuracy at high redshift is
clearly more difficult.

The NIR LF of cluster galaxies at high redshift ($z \geq 0.8$) has
been measured by \citet{trenthamemobasher1998, depropris1999,
nakata2001, kodamaebower, toft2003, ellisejones}, and
\citet{toft2004}. The evolution of the characteristic magnitude
$M^{*}$ was first studied by \citet{depropris1999} from low redshift
up to $z\simeq 0.9$, finding that it is consistent with pure
luminosity evolution of a stellar population formed at $z >2$; this
result has been confirmed by subsequent studies. The evolution of the
faint--end slope was only studied by \citet{toft2003} and
\citet{toft2004}, who found a flatter slope at higher redshift
compared to the local value.

The adopted cosmology in this paper is H$_{0} = 70$ km s$^{-1}$ Mpc$^{-1}$, $\Omega_{M}=0.3$, $
\Omega_{\Lambda}=0.7$ unless otherwise stated. Magnitudes are in the AB system.


\section{Data}

This work is mainly based on near--infrared images of three distant
galaxy clusters: RDCS J1252.9-2927 at $z=1.24$ (hereafter Cl1252,
\citet{rosati2004}), RX J0848+4453 at $z=1.27$ (hereafter Cl0848,
\citet{stanford1997}), and RDCSJ0910+5422 at $z=1.11$ (hereafter Cl0910,
\citet{stanford2002}).  The main properties of these three clusters are
listed in table \ref{tab:properties}. As the data used in this work
have already been published, we refer the reader to the papers listed in table
\ref{tab:data} for details.

\begin{table*}
\centering
\caption{Main properties of the cluster sample. Data are from \citet{ettori2004}. \label{tab:properties}}
    \begin{tabular}{@{}l l l l l l@{}}
\hline
Cluster&z&T$_{gas}$&R$_{500}$&L$_{bol}$&M$_{tot}$\vspace{0.06cm}\\
 & &keV&kpc&$10^{44} erg s^{-1}$&$10^{14} M_{\odot}$\vspace{0.06cm}\\
\hline
RDCS J0910+5422& 1.106&$6.6^{+1.7}_{-1.3}$&$818 \pm 150$&$2.83 \pm 0.35$&$4.91 \pm 2.93$\vspace{0.06cm}\\
RDCS J1252.9-2927&1.237&$5.2^{+0.7}_{-0.7}$&$532 \pm 40$&$5.99 \pm 1.10$&$1.59 \pm 0.35$\vspace{0.06cm}\\
RX J0848+4453&1.273&$2.9^{+0.8}_{-0.8}$&$499 \pm 115$&$ 1.04 \pm 0.73$&$1.37 \pm 0.98$\vspace{0.06cm}\\
\hline
     \end{tabular}
\end{table*}

We used a K$_{s}$ band image of the Cl1252 field obtained with the
ISAAC infrared imager at the VLT \citep{lidman2004}, a K$_{s}$ band
image of the Cl0910 field obtained with the Prime--Focus Infrared
Camera at the Palomar 5m telescope \citep{stanford2002}, and a F160W
($\simeq$ H band) image of the Cl0848 field obtained with the NICMOS
Camera 3 on the Hubble Space Telescope. While the quality of the two
images for Cl1252 and Cl0848 is excellent 
(the PSF has FWHM $\simeq 0\farcs45$ and $0\farcs22$ respectively, 
with limiting AB magnitude $\approx 25$), 
the Cl0910 image has relatively poorer quality (FWHM
$\simeq 0\farcs9$). While for both Cl1252 and Cl0910 the image
effectively used has a radius $\simeq$ 65'' (i.e. slightly more than
500 kpc in linear scale), the NICMOS image for Cl0848 is relatively
small (the maximum radius of the mosaic is $\simeq$ 55'',
i.e. $\simeq$ 450 kpc at z=1.27).  For all the images a catalog was
produced with the SExtractor software \citep{sextractor}, and {\textsc
MAG\_AUTO} was used as a measure of the total magnitude.

\begin{table*}
\centering
\caption{Summary of the principal characteristics of the NIR images used for the determination of the LF. Area$_{eff}$ is the area of the region actually used in this work. \label{tab:data}}
    \begin{tabular}{@{}l l l l l l l@{}}
\hline
Cluster&Telescope/Instrument&Filter&resolution&completeness&Area$_{eff}$&references\vspace{0.06cm}\\
 & & &$\farcs$&AB mag&arcmin$^{2}$ / Mpc$^{2}$& \vspace{0.06cm}\\
\hline
RDCS J0910+5422&Palomar 5m / PFIC&K$_{s}$&$0\farcs9$&21.5&4.35 / 1.06&\citet{stanford2002}\vspace{0.06cm}\\
RDCS J1252.9-2927&VLT / ISAAC&K$_{s}$&$0\farcs45$&24.5&3.69 / 0.93&\citet{rosati2004}\vspace{0.06cm}\\
& & & & & & \citet{lidman2004}\vspace{0.06cm}\\
& & & & & & Demarco et al. (in prep.)\vspace{0.06cm}\\
RX J0848+4453&HST / NICMOS&F160W&$0\farcs22$&25&1.84 / 0.47&\citet{stanford1997}\vspace{0.06cm}\\
& & & & & & \citet{holden2004}\vspace{0.06cm}\\
& & & & & & \citet{vandokkum2003}\vspace{0.06cm}\\
\hline
     \end{tabular}
\end{table*}

\begin{figure}
\centering
\includegraphics[width=8.8cm]{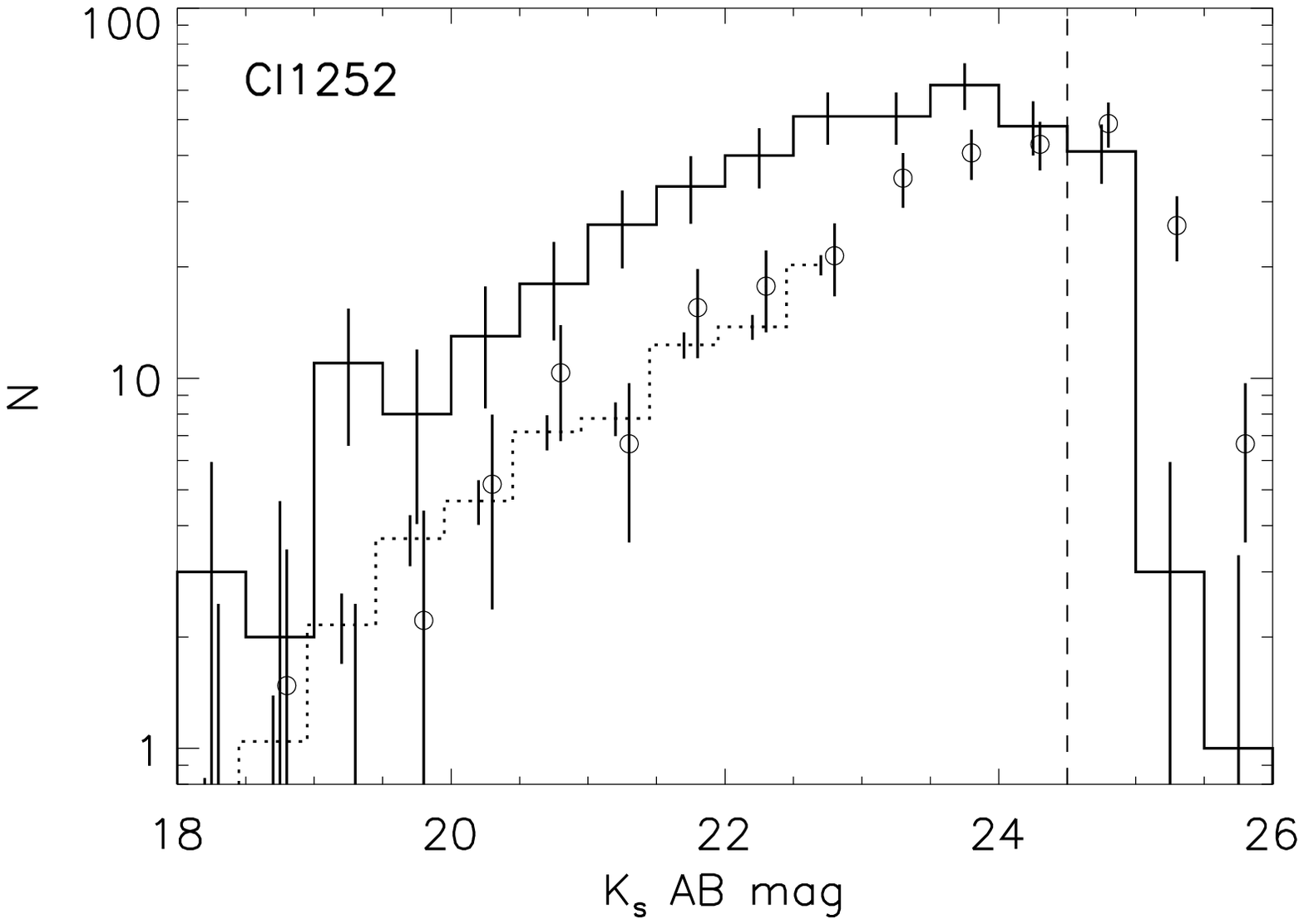}
\includegraphics[width=8.8cm]{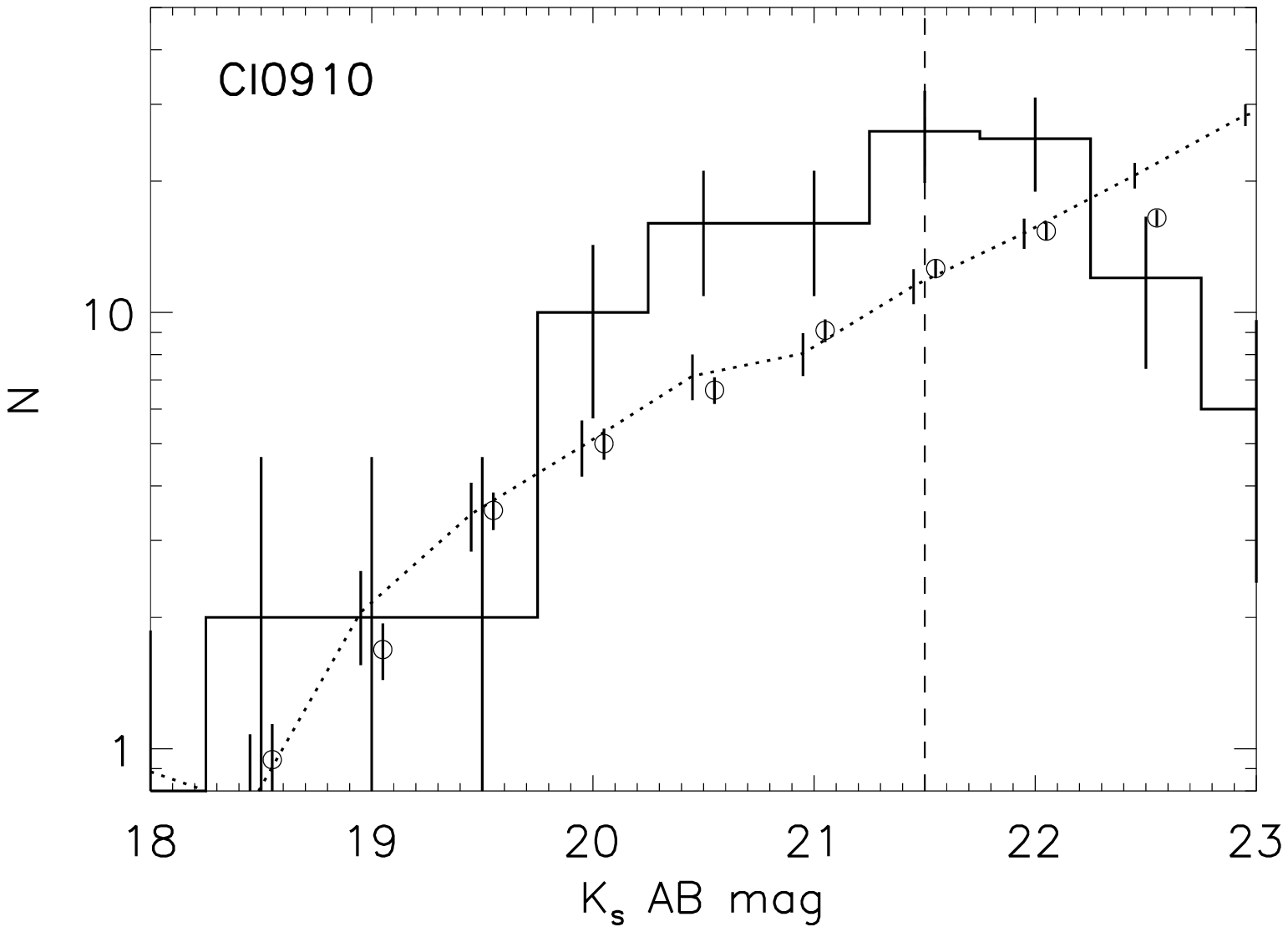}
\includegraphics[width=8.8cm]{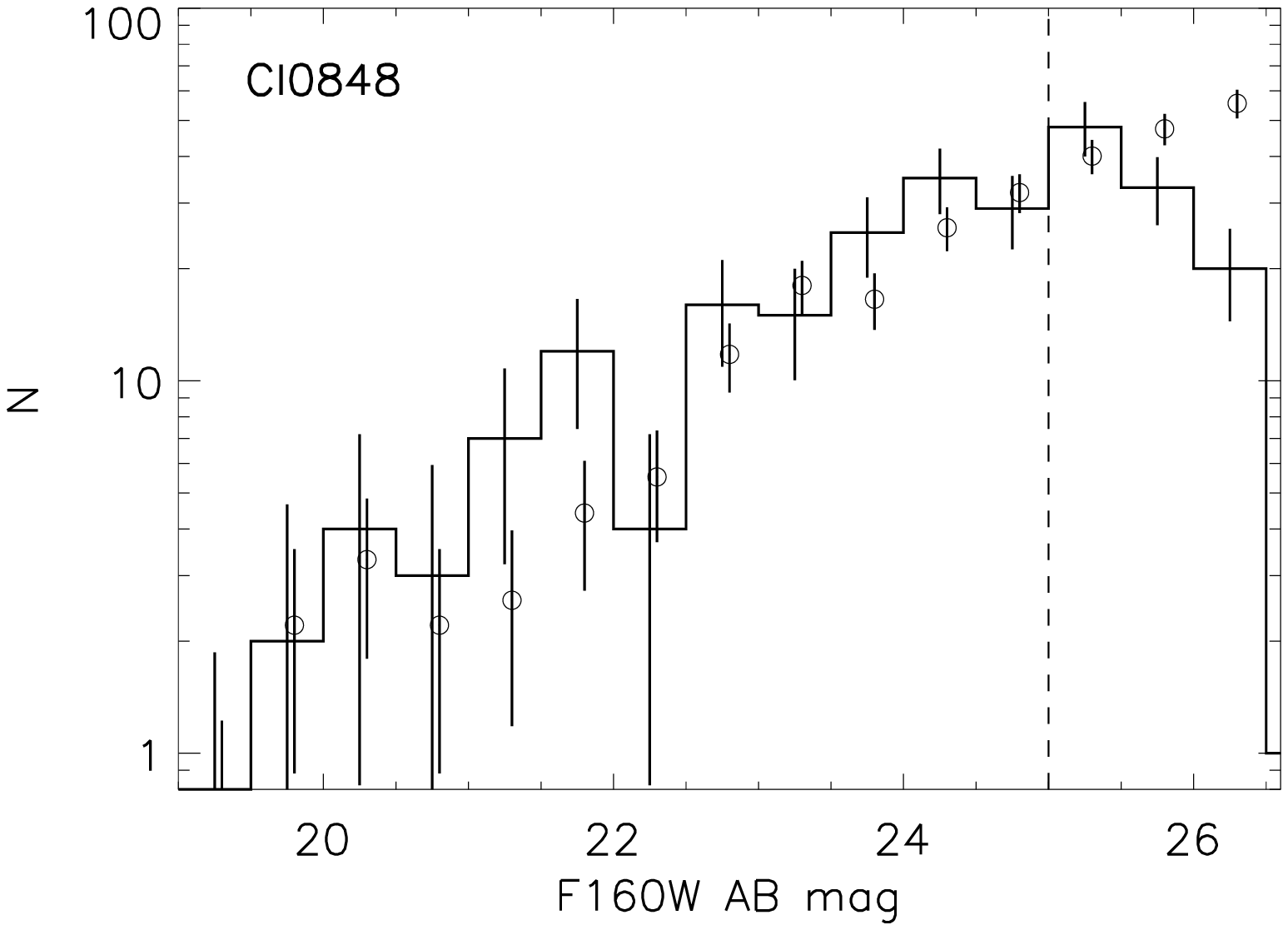}

\caption{Number counts of extended objects with S/N $>$ 5 in the clusters
and reference fields. Upper panel: the solid line shows the number
counts in the Cl1252 field ($\simeq 3.7$ square arcmin). The circles
and the dotted line show the counts in the reference regions selected
in the FIRES and GOODS--S fields respectively, normalized to the cluster area. The
counts in the GOODS--S field are shown down to the completeness
magnitude (K$_{s} \simeq 23$). Middle panel: the solid line shows the
number counts in the Cl0910 field ($\simeq 4.4$ square arcmin). The
circles and the dotted line show the expected background level from
reference regions selected in the SOFI and ISAAC GOODS--S fields,
respectively. Lower panel: the solid line shows the number counts in
the Cl0848 field ($\simeq 1.8$ square arcmin), the circles show the
expected background level based on the Hubble UDF. In all panels,
the symbols/lines showing background
number counts have been shifted by $\pm 0.05$ mag along the x--axis to 
avoid overlapping errorbars.
\label{fig:cathistos}}
\end{figure}

As discussed later, the galaxy luminosity function for all three clusters was
determined by means of statistical subtraction of the fore-- and 
background (hereafter background)  contribution. Since the images are too small
to estimate the local background from the images themselves, a control
field was selected for each of the cluster fields in order to
determine the background contribution to the galaxy counts. Ideally,
the control field should be observed in the same filter and in very
similar conditions and depths. For Cl1252 the control field has been chosen in
the FIRES (Faint Infrared Extragalactic Survey,
\citet{franx2000,labbe2003}) field in the HDF--S region, imaged with
the same instrument and in the same filter as the cluster field. Because of its small area, we have complemented
this field at bright magnitudes with a field in the GOODS--S
region, also observed with VLT/ISAAC (Vandame et al, in
preparation). For Cl0848 the reference field has been taken in the
Hubble Ultra Deep Field \citep{thompson2005}, also imaged with the
same instrument and in the same filter as the Cl0848 field. For
Cl0910 we had no control field available imaged with the same
instrument, we then selected the reference field in the GOODS--S Ks images
observed with both NTT/SOFI and VLT/ISAAC. Due to the similarity of
the K$_{s}$ band filters used for the GOODS and Cl0910 images, we
expect the background estimate to be appropriate. In particular, the
SOFI image has comparable seeing ($\simeq 0\farcs9$) and comparable
completeness magnitude (see figure \ref{fig:cathistos}), and has the
further advantage of being wider (smaller Poissonian errors).

For the purpose of identifying point--like sources we made use of the
HST/ACS images available for all the cluster fields
\citep{blakeslee2003,postman2005} and for the reference regions in
GOODS--S \citep{giavalisco2004} and HUDF (Beckwith et al., in
preparation), and of the HST/WFPC2 for the reference region in the
HDF--S field \citep{williams2000}. In all catalogs, whenever possible
the point--like sources were removed based on the {\textsc MAG\_AUTO}
vs. {\textsc FLUX\_RADIUS} plot derived from the ACS images. The removed
sources have {\textsc FWHM} close to the PSF of the image. 
Point--like sources in regions not covered by the ACS image were
identified in the SOFI image itself. Since point--like sources are
a small fraction of the total counts, uncertainties in their removal
have little effect on our results.

In order to estimate the luminosity function down to the faintest
magnitude allowed by the data, the reference field has to be complete
at least down to the completeness magnitude of the cluster field. In
figure \ref{fig:cathistos} we plot the number counts in the cluster
and reference fields (normalized to the cluster field area). In all
cases the completeness magnitude of the reference field is fainter or
similar to that of the cluster field, therefore the following analysis is
based on the cluster and reference fields catalogs down to the cluster
field completeness magnitude, without completeness corrections.  From
the turn--over of the number counts for objects with $S/N$ ratio $> 5$,
the completeness magnitude of the cluster fields is estimated to be
K$_{s}$ = 24.5 for Cl1252, K$_{s}$ = 21.5 for Cl0910, and F160W = 25
for Cl0848.

This work also makes use of the extensive spectroscopic campaigns in
these three clusters. We refer to
\citet{stanford1997,stanford2002,vandokkum2003} and Demarco et al. (in
preparation) for details on the spectroscopic follow--up observations.

\section{Luminosity functions}

For each of the clusters the luminosity function (LF) was calculated
by means of statistical subtraction, i.e. using a reference field to
remove the background contribution in the assumption that the field
galaxy density is constant all over the sky. The statistical
subtraction of the field galaxies is often considered to be an
uncertain method of background removal especially at high redshifts,
where the signal of the cluster against the background is
progressively lower. However, while obtaining spectroscopic redshifts
for all cluster galaxies down to a reasonably faint magnitude is
clearly unfeasible, even the determination of photometric redshifts
relies on the availability of deep photometry in several passbands,
and on the assumption that the photometric redshift quality remains
the same for spectral energy distributions (SED) for which no
spectroscopic redshift can be measured.

Only one (Cl1252) of the three clusters studied has such a deep and
wide photometric coverage (in addition to 38 spectroscopically
confirmed members), that cluster membership determination fully
based on photometric redshifts is feasible; the LF for Cl1252 was determined in
this way by \citet{toft2004} (hereafter T04). Since for the other two
clusters statistical subtraction is at present the only viable
approach, the LF of Cl1252 was re--determined with this method, in the
same region as in T04, as a first--order validation of the statistical
subtraction procedure in this redshift range.

\subsection{LF determination}

For each cluster the LF was determined as follows.  The galaxy counts
in both the cluster and reference fields were binned, and the
background contribution was estimated in each bin as the reference
field counts normalized to the cluster area. For each bin, the error is
estimated as the sum in quadrature of the Poissonian errors $1 +(N
+0.75)^{1/2}$ \citep{gehrels} on both cluster and
field counts in that bin.

The error on the background counts should also take into account the
effects of galaxy clustering and of the lensing magnification of
galaxies beyond the cluster. However, an estimate of the galaxy
clustering contribution to the number counts error according to the
prescriptions of \citet{huang1997} yields a negligible difference with
respect to the simple Poissonian error.  Since a significant fraction
of the observed galaxies are in the foreground of the cluster the
effect due to lensing is likely small. Due to the large Poissonian
errors we have, we can neglect these effects.

The existing spectroscopic data were taken into account, so that in each
bin the background contribution contained at least as many galaxies as 
the spectroscopic interlopers, and the background corrected counts
were at least equal to the number of confirmed members. This allows a
more secure determination of the LF in the bright end, where due to
low counts (both in the cluster and in the control fields) the
statistical subtraction may be poor. In the area selected for the LF
determination, most of the bright galaxies have measured redshifts:
almost 80\% of the galaxies down to Ks = 21 for Cl1252, more than 80\%
down to Ks = 21 for Cl0910, and more than 85\% down to F160W = 22 for
Cl0848.

Since the FIRES field is quite small, in order to achieve a better
background evaluation (and smaller errors) in the bright end, the background
estimate for Cl1252 from the FIRES field was supplemented with the
estimate from the control field in GOODS--S (ISAAC) for magnitudes
brighter than 21.5.

Due to the lower quality of the Cl0910 K$_{s}$ image (seeing $\simeq
0\farcs9$), special care was taken for blended objects,
particularly in the overdense cluster environment. Thanks to the
availability of HST/ACS images in passbands F775W and F850LP, it was
possible to crosscheck the catalogs to identify obvious
blendings. Eight cases of evident blending were identified: for six of
them, a more 'aggressive' SExtractor configuration allowed the blended
source to be split in sources located as in the ACS images. For the
remaining 2 cases no deblending could be achieved, and as a zero--order
approximation the flux from the source was split according to the flux
ratio of the blended sources in the HST/ACS F850LP image.

\begin{figure}
\centering
\includegraphics[width=8.8cm]{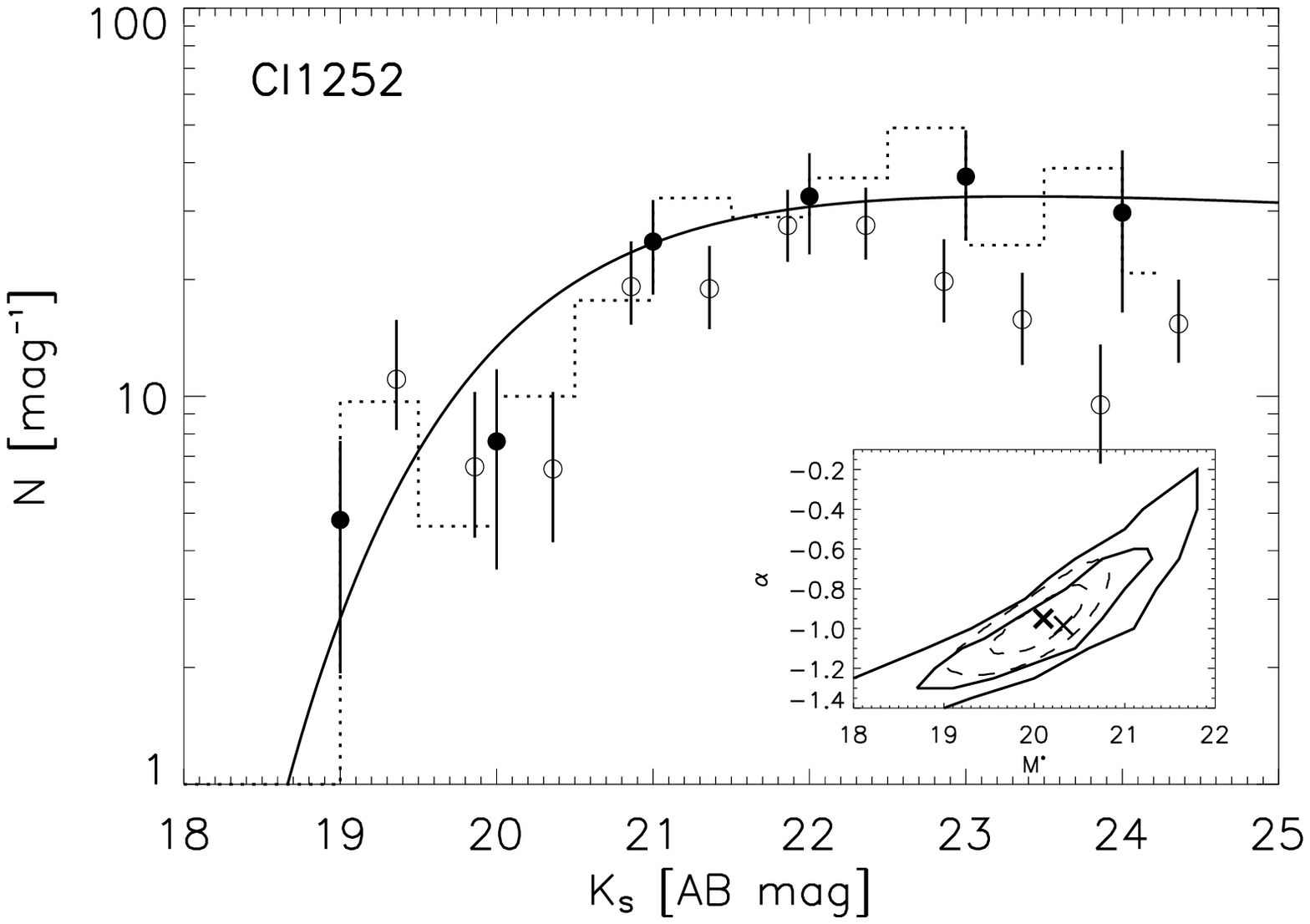}
\includegraphics[width=8.8cm]{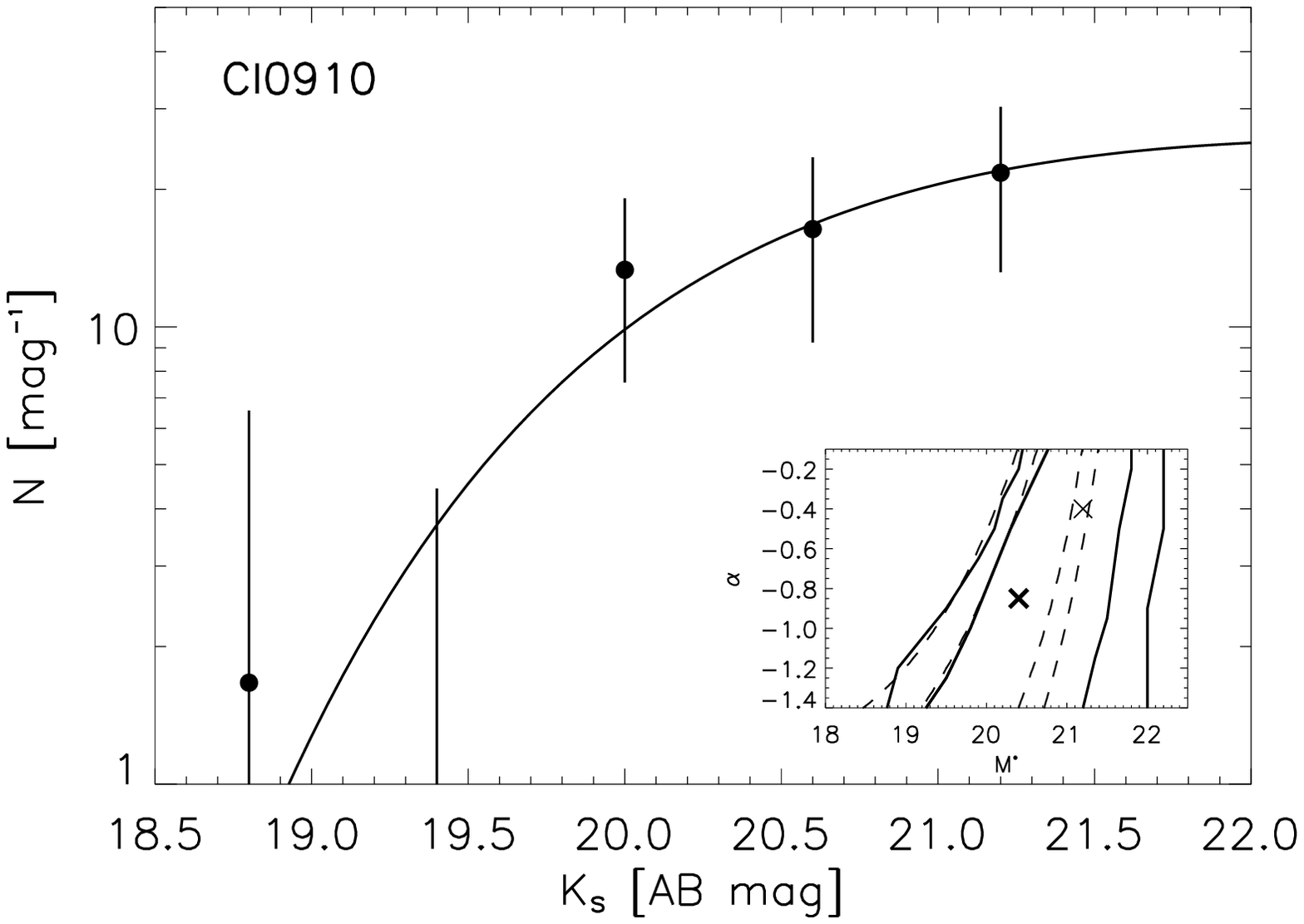}
\includegraphics[width=8.8cm]{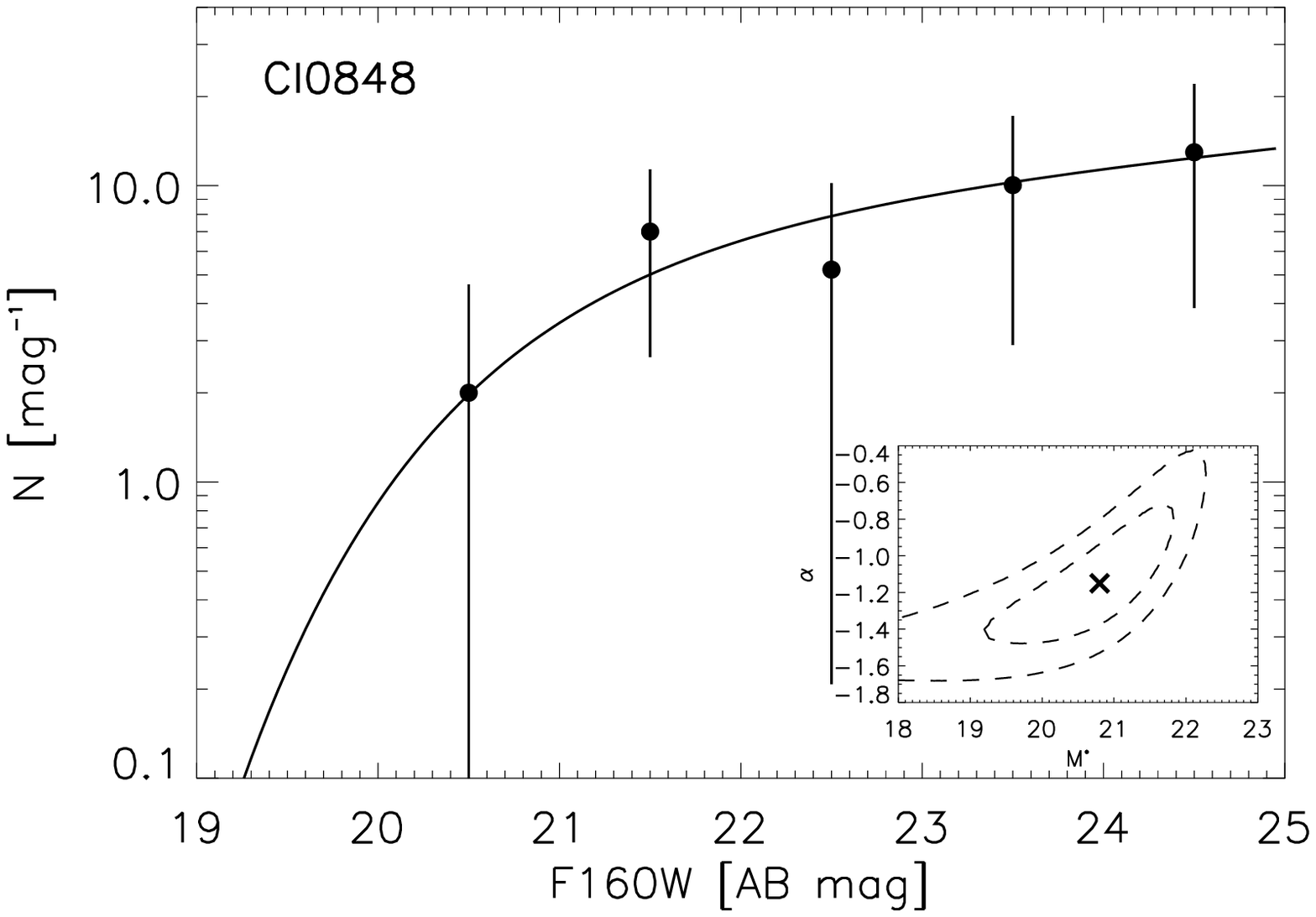}

\caption{Individual cluster luminosity functions for Cl1252 (upper
panel), Cl0910 (middle), and Cl0848 (lower). In all panels, the solid
line shows the best--fit Schechter function as derived from the binned
counts, and the inserted plot shows the 1, 2--$\sigma$ confidence
levels on the parameters M$^{*}$ and $\alpha$, calculated from the
maximum likelihood on binned counts (dashed lines) and from the
maximum likelihood on unbinned, not background subtracted counts
(solid thick lines -- not shown for Cl0848, see text).  The thick and
thin crosses show the position of the best--fit parameters derived
with the former and latter methods, respectively (see text for
details).  In the upper panel the empty circles show the Cl1252 LF as
determined by T04, and the dotted histogram shows for comparison the
LF from this work binned with the same bin size as in T04.
\label{fig:LFclusters}}
\end{figure}

The luminosity functions are shown in figure \ref{fig:LFclusters}.
The binned LFs were fit with the usual \citet{schechter} function with
a maximum likelihood method using the \citet{cash} statistics $C =
-2\Sigma_{i}[n_{i}$ ln$(m_{i}) -m_{i} -$ ln$(n_{i}!)]$, where $n_{i}$
and $m_{i}$ are the observed number of galaxies in the $i^{th}$
magnitude bin, and the number of galaxies predicted in the same bin by
a Schechter function of parameters M$^{*}$ and $\alpha$,
respectively. The best--fit M$^{*}$ and $\alpha$ are the parameters
that minimize $C$. $\Phi^{*}$ was not taken as a free parameter, but
was calculated for each choice of M$^{*}$ and $\alpha$ by requiring
that the total number of predicted galaxies equal the number of those
actually observed.

Even though the faint end slope cannot be well constrained (or is
completely unconstrained, as in the case of Cl0910), due to the well
known correlation of the Schechter parameters, leaving both M$^{*}$ and
$\alpha$ free  allows a better evaluation of the errors on M$^{*}$.
The best--fit Schechter functions are overplotted on the LFs in figure
\ref{fig:LFclusters}.  The best fitting Schechter parameters are
listed in table \ref{tab:LFs}.

The maximum likelihood approach gives in principle also an estimate of
the confidence levels on the best--fit parameters, as if the
Cash statistics is defined as above, $\Delta C$ is distributed like
$\Delta \chi ^{2}$, thus $\Delta C = 2.3, 6.17$ gives the 1,
2--$\sigma$ confidence levels for two interesting parameters M$^{*}$
and $\alpha$.  However, it should be noted that the Cash
statistics should be applied to data which include background, because
the background subtracted data are not
Poisson--distributed,
while the Cash statistics  assumes
Poisson probabilities.
Even if we believe that the relevance of the
80\% spectroscopic completeness on the LF bright end is important
enough to adopt the previously described approach on binned, background subtracted counts, we note that in our LF fitting
approach we hid the fact that the counts in each bin have errors
larger than Poissonian due to the previous statistical background
subtraction, and therefore we tend to underestimate the errors on
M$^{*}$ and $\alpha$. For this reason, we also adopt a maximum
likelihood approach on unbinned, not background subtracted
data. Recently \citet{andreon2005} summarized the principal reasons
why one should adopt this  
approach, and proposed a method to be applied when the individual
membership of the galaxies is unknown. When applying this method we
are thus neglecting our redshift information, which means that we will
derive conservative confidence intervals. 

For each of the three clusters we applied the method described in
\citet{andreon2005}, taking as background dataset both the control
regions in GOODS(ISAAC) and FIRES for Cl1252, and the control regions
in GOODS(SOFI) and UDF for Cl0910 and Cl0848 respectively.  In brief,
we assume that the background number counts can be described by a
power law (we used three parameters), and that the cluster LF is a
Schechter function (also three parameters: M$^{*}$, $\alpha$, and
$\Phi^{*}$). We then find the parameters that at the same time
maximize the likelihood for the number counts in the control field
(only described by the power law), and in the cluster field (described
by the power law plus the Schechter function).  In all cases, we first
searched the complete 6 parameter space for the global maximum, and
then found the maximum likelihood on a grid in the M$^{*}$--$\alpha$
plane (i.e. varying only the remaining four parameters), so that we
can draw the confidence levels for these two parameters. Due to very
low counts (as discussed below), 
it was difficult to maximize
the likelihood against M$^{*}$ and $\alpha$ for Cl0848. Since the
constraints on the Schechter parameters as determined with the first
approach are already very loose, for this cluster we quote them in the
following.

In figure \ref{fig:LFclusters} we show for reference the 1-- and
2--$\sigma$ confidence levels obtained with the two different
approaches for Cl1252 and Cl0910. As we mentioned, the smaller ones
are understimated but the larger ones are likely overestimated, thus
the `real' confidence levels are expected to lie between the two.  The
errors we quote in the following for Cl1252 and Cl0910 are derived
from the larger ones.

The Cl1252 LF is shown together with the LF based on photometric
redshifts from T04: the two determinations agree within the errors,
with a larger discrepancy for the last magnitude bin. The
difference between the two LFs could be considered a measure of the
systematics of the two methods, with the error budget still dominated
by low number statistics.  The slope $\alpha$ determined via
statistical subtraction ($-0.9 \pm 0.3$) is somewhat steeper than the
one determined via photometric redshifts ($\alpha =
-0.64^{+0.27}_{-0.25}$), however the two estimates are consistent
within the errors.
The K$_{s}^{*}$ value
($20.1^{+1.1}_{-1.2}$) is also found in agreement with the determination by T04
(K$_{s}^{*} = 20.41^{+0.45}_{-0.55}$).

The Cl0910 Schechter function plotted in figure \ref{fig:LFclusters}
is the formal best fit with K$_{s}^{*} = 20.4$ and $\alpha =
-0.85$. As it is clear from figure \ref{fig:LFclusters}, the slope
$\alpha$ is unconstrained. Assuming $-1.4 < \alpha < 0.4$ yields an
error of $\pm 1$ mag on K$_{s}^{*}$.

Finally, some caveats apply to the determination of the Cl0848
LF. Namely, i) the very small field of view (less than 200 objects
brighter than H$_{F160W}$ $\simeq 25$), resulting in significant
Poissonian errors, ii) the likely intrinsically lower richness of
Cl0848 compared to Cl0910 and Cl1252 (since Cl0848 has lower mass),
iii) the presence of a known underlying supercluster in the Lynx
field, and of a lower redshift cluster projected in the supercluster
region \citep{stanford2001, nakata2005}.  Mainly due to the first two
reasons, the constrains on the Schechter parameters are quite loose,
despite the significant depth of the F160W image. The formal best--fit
Schechter parameters are F160W$^{*}$ = $20.8^{+1.}_{-1.6}$ and $\alpha
= -1.15^{+0.4}_{-0.3}$.

\subsubsection{Composite luminosity function}

Summing up the galaxy number counts of different clusters allows the
background subtraction to be more effective (averaging over
uncertainties in the statistical subtraction in each single cluster), and the shot
noise to be reduced.  The composite luminosity function was calculated
in the observed K$_{s}$ band at $z = 1.2$. For this reason the F160W
magnitudes at z=1.27 were k--corrected to K$_{s}$ magnitudes at $z =1.2$.

A single k--correction of 0.5 was applied at all magnitudes, as
derived from synthetic SEDs \citep{bruzualecharlot} of evolved simple
stellar populations at that redshift ($3 \leq $ age $\leq 5$ Gyr)
No correction was made for the
negligible k--correction between $z=1.2$ and $z=1.106$ or $z=1.237$.
All the individual LFs were binned with a bin size of 0.5 mag (binning
was adjusted to optimize individual magnitude coverage taking into
account completeness limits and corrections for different redshift).
Cl1252 and Cl0848 have the same ``K$_{s}$'' band completeness after
these corrections, so their LFs were just summed up, and the errors were
added in quadrature. Cl0910 instead is much shallower, so its LF was
added to the composite LF up to its completeness magnitude. The
composite LF beyond this magnitude is computed as the composite LF
without Cl0910 multiplied by the ratio of the total counts (including
all three clusters) to the total counts excluding Cl0910, computed
in the magnitude interval where the Cl0910 photometry is complete (and
errors were scaled accordingly).  Due to the  bright completeness
limit for Cl0910, and the low counts for Cl0848, it is clear that the
faint end of the composite LF is dominated by Cl1252.  In our case,
building the composite LF following other common methods as described
in \citet{colless} or in \citet{garilli}, produces results consistent
within 1--$\sigma$ with the LF calculated as above (note that the
method described in \citet{garilli} tends to give smaller errors, and
a flatter slope than the one in \citet{colless}).

The composite LF was also derived with a maximum likelihood approach
on unbinned, not background subtracted data, as described in
\citet{andreon2005}. However, since the LF of Cl0848 is measured in a
different passband than those of Cl1252 and Cl0910, Cl0848 was not
included in this case.  The composite LF was determined by fitting at
the same time the counts in the two cluster fields (Cl1252 and Cl0910)
and in all the K$_{s}$ band control fields. The two cluster LFs are
assumed to be described by the same $M^{*}$ and $\alpha$ (but clearly
have two different $\Phi^{*}$).

\begin{figure}
\includegraphics[width=8.8cm]{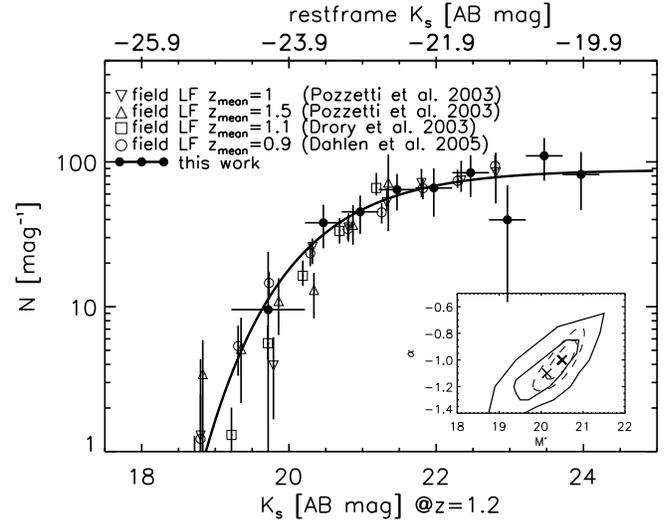}
\caption{The composite cluster luminosity function at z=1.2 (filled
dots) compared to the field galaxies
LF at $z \simeq 1$. The solid line shows the best--fit
Schechter, and the inserted plot shows the 1, 2--$\sigma$confidence
levels on the Schechter parameters as in figure
\ref{fig:LFclusters}. All field galaxies LFs have been
arbitrarily rescaled.
\label{fig:compLFfield}}
\end{figure}

\begin{figure}
\includegraphics[width=8.8cm]{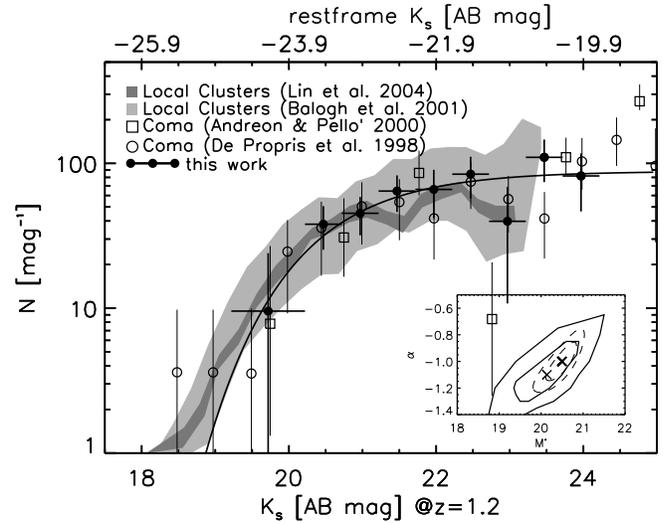}
\caption{The composite cluster luminosity function at z=1.2 (filled
dots)  compared to the
local cluster galaxies LF, corrected by 1.3 mag for passive evolution
(see text). The solid line shows the best--fit
Schechter, and the inserted plot shows the 1, 2--$\sigma$confidence
levels on the Schechter parameters as in figure
\ref{fig:LFclusters}. All local cluster galaxies LFs have been
arbitrarily rescaled.
\label{fig:compLFlocal}}
\end{figure}

The composite luminosity function is shown in figures
\ref{fig:compLFfield} and \ref{fig:compLFlocal}. The best--fit Schechter has K$_{s}^{*} =
20.5^{+0.4}_{-1}$ and $\alpha = -1.0^{+0.2}_{-0.3}$. In figure \ref{fig:compLFfield}
we also show the {\it field} LF at z $\simeq$ 1 from
\citet{pozzetti2003,drory2003} and \citet{dahlen2005}.  In figure \ref{fig:compLFlocal}
the composite LF is compared to the local cluster
galaxies $K$ band LF (corrected by 1.3 mag for passive evolution as
derived below) as measured in Coma by \citet{depropris1998} and
\citet{andreonepello}, and in samples of nearby clusters by
\citet{balogh2001} and \citet{lin2004}. All the local and field LFs
have been arbitrarily rescaled.  Both the Coma LFs shown were measured
in the H band and shifted to K band with a colour term $H-K = 0.24$.
The K band field LFs from \citet{pozzetti2003} and \citet{drory2003},
and the J band field LF from \citet{dahlen2005}, were converted to
observed K$_{s}$ magnitudes at $z=1.2$ by equations 2 and 1 in
\citet{pozzetti2003}.

With these data, no significant difference can be seen between the
shapes of the cluster and field luminosity functions in the probed magnitude
range, even if we find some evidence of an excess of very bright galaxies
with respect to the field, as suggested for instance in
\citet{depropris2003}.

The derivation of luminosity functions based on statistical
subtraction may be affected by field-to-field variations, as the
background evaluated from a control field may not be representative of
the background in the cluster field. A robust estimate of such an
effect requires adequately deep and large K band fields. The
VLT/ISAAC observations of the GOODS--S field, covering $\sim\! 
100\,{\rm arcmin}^2$, are currently the best data set available for
this purpose. We note that part of this same field has been used as control
field for determining the LFs of Cl0910 and Cl1252. In each of 17
selected ISAAC tiles in GOODS--S, we considered all galaxies within the
central region (with the same area as the Cl1252 field used for the
LF), down to the photometric completeness.  We also considered in this
case as an ``additional tile'' the FIRES HDF--S field we already used
as control field for Cl1252.

\begin{figure}
\centering
\includegraphics[width=8.8cm]{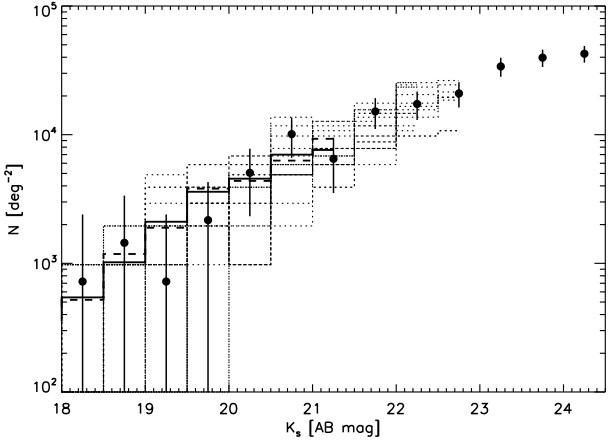}
\caption{The background number counts as estimated in different
control fields. The filled symbols show the counts from the control
field in the FIRES region, the solid and dashed lines from those in
the GOODS--S regions (ISAAC and SOFI, respectively), and the dotted
lines show the counts from the 17 ISAAC tiles in the GOODS--S field.
\label{fig:backcounts}}
\end{figure}

In figure \ref{fig:backcounts}, we
show the background number counts as estimated from different control
fields: those from the FIRES field, which was used as control
field for Cl1252 (area $\simeq$ 5 arcmin$^{2}$) at faint magnitudes,
those from the ISAAC GOODS--S mosaic used to complement the Cl1252
control field at bright magnitudes (area $\simeq$ 53 arcmin$^{2}$),
those from the SOFI GOODS--S mosaic used for Cl0910 (area $\simeq$ 152
arcmin$^{2}$), and those from the 17 ISAAC tiles in GOODS--S (area $\simeq$ 3.7
arcmin$^{2}$ each).

We then built the Cl0910 and Cl1252 LFs as described above
(making use of the spectroscopic information), based on these 18 small
control fields, and for each of these we redetermined the composite
LF. For Cl0848 we always used the LF determined above, since due to
its very large uncertainties it has lower weight compared to Cl0910
and Cl1252.  Thus, we obtained 18 composite LFs corresponding to
different control fields.

The GOODS--S ISAAC data, while being considerably wide, are not deep
enough to reach our faint--end magnitudes. Therefore, we complemented
the counts from each ISAAC tile at magnitudes fainter than its
completeness with the FIRES counts. In this way, we probe
field-to-field variations at magnitudes brighter than K$_{s}\simeq 23$. The
medians and standard deviations of M$^{*}$ and $\alpha$ obtained from
these 18 determinations are M$^{*}$ = 20.5 mag,
$\sigma_{M^{*}}$=0.14 mag, $\alpha$ = -1.0, $\sigma_{\alpha}$ =0.06.

In order to also account for field-to-field variations at fainter
magnitudes (probed only by FIRES), we repeated the LFs determination by
normalizing the FIRES counts at faint magnitudes by the ratio of the
ISAAC/FIRES counts at magnitudes brighter than the completeness in
each tile, assuming that the number density ratio is the same at
fainter magnitudes.  The medians and standard deviations of M$^{*}$
and $\alpha$ obtained from the 18 redeterminations with this procedure
are M$^{*}$ = 20.5 mag, $\sigma_{M^{*}}$=0.09 mag, $\alpha$ = -1.0,
$\sigma_{\alpha}$ =0.09.

\begin{figure}
\centering
\includegraphics[width=8.8cm]{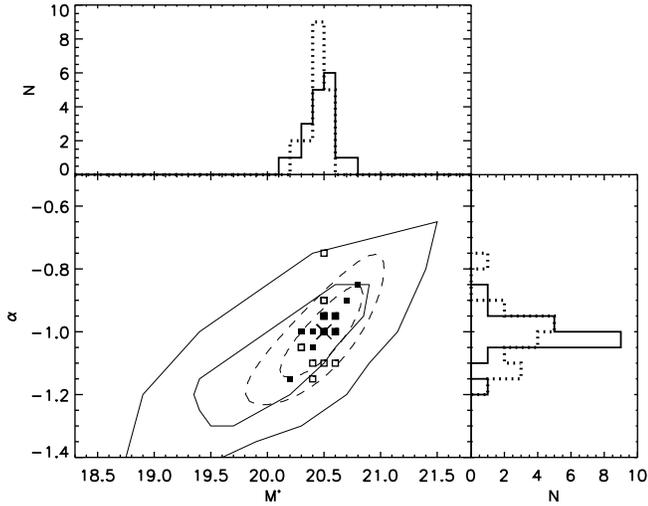}
\caption{Impact of field-to-field variations on the determination of the LF.
In the main panel the best--fit M$^{*}$ and $\alpha$ (cross) and their
1-- and 2--$\sigma$ confidence levels are shown, as in figures
\ref{fig:compLFfield} and \ref{fig:compLFlocal}. The M$^{*}$ and $\alpha$ parameters from LF
determinations with 18 different small control fields are shown as
squares. In the side--panels the distributions of these M$^{*}$ (top) and
$\alpha$ (right) are shown. Solid symbols (in the main panel) and
solid lines (in the side panels) show the results for GOODS--S tiles
counts complemented with FIRES counts at faint magnitudes, while empty
symbols and dotted lines show the results with modified FIRES counts
at faint magnitudes (see text for details).
\label{fig:FFVs}}
\end{figure}

In figure \ref{fig:FFVs}, we show the results from these two sets of
tests. In the main panel, we show the confidence levels as shown in
figures \ref{fig:compLFfield} and \ref{fig:compLFlocal}, and we overplot the Schechter parameters
M$^{*}$ and $\alpha$ obtained for the 18 LFs. Since most of the data points
overlap near the original determination marked by the cross, we show
their distributions in the side--panels.

\subsection{Evolution of the restframe K band LF}

It is customary to compare the observed evolution of M$^{*}$ with
redshift with different predictions.  Even if initially
\citet{barger1998} found no significant evolution for the infrared
M$^{*}$ up to $z>0.5$, first \citet{depropris1999} and then other
works on high redshift clusters
\citep{nakata2001,kodamaebower,massarotti2003,toft2003,ellisejones}
found that the evolution of M$^{*}$ up to $z \simeq 1$ is consistent
with pure luminosity evolution of the cluster galaxies, and
inconsistent with no--evolution predictions. As discussed in
\citet{andreon2001} and \citet{andreon2004}, measuring an evolution in
the LF from a change in the best--fit Schechter parameters is not
straightforward. A density (clustercentric radius) dependent LF would
imply a dependence of M$^{*}$ (and $\alpha$) on the surveyed area in
different clusters at different redshifts, and the correlation of
M$^{*}$ and $\alpha$ could introduce spurious results. 
 This makes it difficult to study the
LF evolution exclusively based on the evolution of the characteristic
magnitude M$^{*}$.  
However, it is unlikely that the bright end of the LF
is dominated by galaxies in the cluster outskirts, and even
though the LF has indeed been found to be dependent on the sampled
region within the cluster, this dependence mostly affects the LF at
magnitudes fainter than those we can probe in this work (e.g.,
\citet{popesso2006}). When comparing LFs from different studies, we
note that our LF is based on the central $r\leq 500$ kpc cluster
regions, approximately corresponding to $r_{500}$, while the
\citet{depropris1998} LF is determined in the central $r \leq 350$ kpc
Coma region, the \citet{andreonepello} LF is determined in a $\simeq
500 \times 500$ kpc$^{2}$ region offset by $\simeq 360$ kpc from the
Coma centre, and the \citet{lin2004} LF is determined within the
virial radius and is found to be very similar to the LF
derived within r$_{500}$.

Finally, if $\alpha$
is free in the Schechter fit, the errors on M$^{*}$ are reliable enough
to make a fair comparison of the M$^{*}$ evolution with different
predictions.

\begin{figure}
\centering
\includegraphics[width=8.8cm]{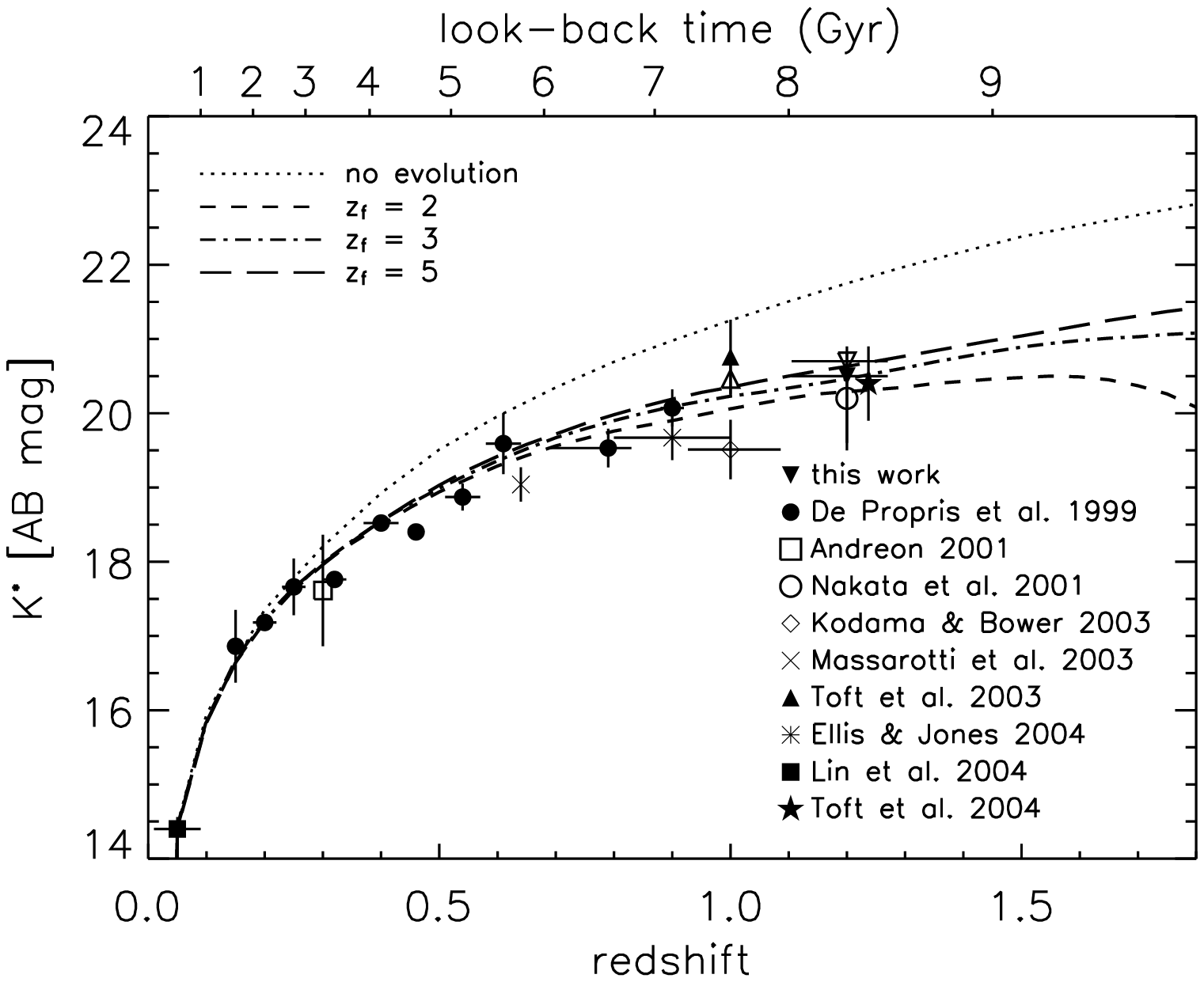}

\caption{The redshift evolution of the characteristic magnitude
K$^{*}$. Different symbols show different determinations of K$^{*}$ as
indicated in the legend. The no evolution prediction is calculated
from the Coma K$^{*}$ \citep{depropris1998} , k--corrected as in
\citet{pozzetti2003}. Passive evolution model predictions are from
\citet{kodamaearimoto}, normalized to the Coma K$^{*}$. All
determinations have fixed slope $\alpha$ = -0.9, except
\citet{andreon2001,toft2003,lin2004}, T04 and the present work. For
\citet{lin2004} the plotted value is the mean of two values determined
for $\alpha = -0.85$ and $\alpha = -1.1$. For both this work and
\citet{toft2003} also the K$^{*}$ value with $\alpha$ fixed at -0.9 is
shown (corresponding empty symbol). Note that errors on M$^{*}$ which
are computed with $\alpha$ fixed and free are not directly
comparable. Error bars on the x--axis, when plotted, represent the
redshift range of clusters which are combined to draw that point.
\label{fig:Kevol}}
\end{figure}

The comparison with previous determinations of K$_{s}^{*}$ at lower
and similar redshift is shown in figure \ref{fig:Kevol}.  In agreement
with previous work, the measured K$_{s}^{*}$ is consistent
with passive evolution predictions for an L$^{*}$ galaxy formed at
z$\geq 2$.

Converting the observed K$_{s}^{*}$ to the absolute K$_{s}$ band
magnitude via K$_{s,rest}$ = K$_{s,obs}$ - 5log(d$_{L}$/10 pc) -
(K$_{s,rest}$ - K$_{s,obs}$)$_{z}$, as in \citet{pozzetti2003}, gives
K$_{s,rest}^{*} = -23.41^{+0.4}_{-1}$. Compared to the Coma LF
K$_{s}^{*} \simeq -22.15$ \citep{depropris1998}, this yields an evolution of $\Delta K^{*}
= -1.3^{+0.5}_{-1}$.  As shown in figure \ref{fig:compLFlocal}, the shape of  the 
composite LF is very similar to the shape of the local cluster galaxies LF shifted 1.3
magnitudes brighter. 

This shape may be quantified in a non--parametric way by an
analogue of the `giant--to--dwarf ratio' (GDR), which is defined in
this case as the ratio of the number of galaxies brighter than
K$_{s}$ = 21.2 to the number of galaxies with $21.2<$ K$_{s} <
24.2$. The K$_{s}$ = 21.2 threshold corresponds to an absolute
magnitude of $\simeq -22.7$, and to a stellar mass of $\simeq 8
\times 10^{10}$ M$_{\odot}$ (for a Salpeter IMF). For our composite
LF this GDR is $0.2 \pm 0.1$.  If we estimate this GDR with the same
absolute magnitude cut using the Coma LFs as shown in figure
\ref{fig:compLFlocal} (i.e. keeping into account the 1.3 mag brightening), we similarly find a GDR
of $0.2 - 0.3$. This suggests that a large fraction of the giant
population was already in place at $z\sim\! 1$.  We remind that a
computation of evolution corrections, k--corrections, and H-K colour
terms at redshift zero, are involved in such a comparison. We also
remind that we are applying a single 1.3 mag brightening for the whole
LF, which is clearly a simplistic assumption, since galaxies with
different star formation histories have different evolution corrections.

\subsection{Contribution to the LF from early and late type galaxies}

The study of the early and late type galaxies LFs at this redshift is
challenging. 

Even when deep ACS data are available,(as in the case of \ Cl1252),
morphological analysis is not feasible at the faint end
(z$_{AB}\gtrsim 25$). 
Here we use the morphological information
presented by \citep{blakeslee2003}, which is based on $i$ or $z$ band
ACS imaging. Due to the red  $B-z$ restframe colours (corresponding
approximately to the observed $z-$K), it is not possible to
build a morphological catalog down to the K$_{s}$ band completeness
magnitude. On the other hand, if early and late types are
distinguished based on their colours, while the cluster red galaxies
can still be isolated with relatively small background pollution, this
does not hold for the blue population.  For these reasons we only
attempt a separation of the contributions to the LF from early and
late type galaxies for the bright end of the Cl1252 LF.

For this cluster we can use the photometric members selected in T04,
and distinguish early and late types based on the best--fitting
template from photometric redshifts. The photometric redshifts in T04
were determined with 7 passbands against four templates (E/S0, Sbc,
Scd, Irr) from \citet{coleman1980}, two starburst SEDs from
\citet{kinney1996}, and interpolations between these six SEDs (see T04
for details). We then defined as early--types those galaxies best
fitted with SEDs earlier than midway between E/S0 and Sbc
(i.e. roughly including E/S0 and possibly some Sa galaxies).  We can
then separate the two contributions to the LF, virtually down to the
K$_{s}$ band completeness magnitude, using the completeness function
calculated by T04 (their fig. 5) to correct both the early and
late--types LFs for incompleteness due to photometric
redshifts. 
While the reliability of photometric redshifts may be uncertain at the
faint end, the early/late-type separation at the bright end is robust
and the completeness correction negligible.

\begin{figure}
\centering
\includegraphics[width=8.8cm]{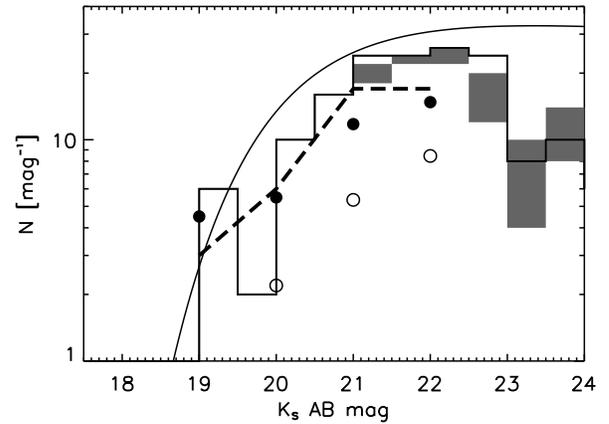}
\caption{The contribution of early and late type galaxies to the
  LF of Cl1252. The filled and empty symbols show the LFs for early
  and late type photometric members (based on the T04 sample),
  classified from their broad band colours. The dashed line shows the
  contribution of morphologically selected early--types
  \citep{blakeslee2003}. The solid histogram shows the number counts
  of all galaxies (excluding spectroscopic interlopers) along the red
  sequence as determined in \citep{blakeslee2003} (no statistical
    subtraction of the field galaxies contamination was made; the
  shaded area shows the effect of photometric errors -- see text). The
  solid curve shows the best--fit Schechter function for the Cl1252
  global LF.
\label{fig:LFbytype}}
\end{figure}

In figure \ref{fig:LFbytype} the early (filled dots) and late--types
(empty circles) bright end LFs are shown, where the separation in
early and late--types reflects their SED properties.  We also show as
a dashed line the LF of early type galaxies morphologically selected
by \citet{blakeslee2003} (brighter than K$_{s} <22.5$, as the typical z-K
colours drive the sample beyond the z$_{850}$ completeness limit at
fainter magnitudes).  The morphologically and SED--selected
early--types LFs are in very good agreement, suggesting that the
bright end of the LF is already dominated by early--type galaxies,
either selected on their morphology or on their spectrophotometric
properties.
 
 A histogram of the red--sequence galaxies (which are expected to be
mostly early types) is also shown for comparison This was derived
taking all the galaxies (within 65\farcs \ from the cluster center)
with i$_{775}$-z$_{850}$ colours 0.16 mag redder or 0.14 mag bluer
than the red sequence determined by \citet{blakeslee2003}. This is
much larger than the intrinsic scatter found by \citet{blakeslee2003},
however the colours we used are 1$\farcs$5 aperture colours,
correspondingly the scatter is expected to be larger.
 No statistical subtraction was attempted in this case, since a
reference field with deep enough K$_{s}$, i$_{775}$ and z$_{850}$
imaging is available, however spectroscopic interlopers were
removed. As a result, the histogram shown is an upper limit to the
effective LF of red sequence galaxies. The shaded area represents the
16--84 percentile variations of this histogram due to photometric
errors, and was derived by simulating 100 catalogs where the
i$_{775}$-z$_{850}$ colour was randomly shifted within a Gaussian of
$\sigma$ equal to the photometric error on the i$_{775}$ - z$_{850}$
colour.  This histogram also confirms that the LF bright end is
largely dominated by galaxies hosting evolved stellar populations.

A solid determination of the LF of red-sequence galaxies would
require even more extensive redshift information at the faint end,
which is however beyond the current spectroscopic limit. With our
data, we observe some evidence of a deficit of faint galaxies on the
red--sequence, which has been reported in other studies
(\citet{kajisawa2000,nakata2001,kodama2004,delucia2004,tanaka2005}),
and is usually interpreted as a sign of downsizing.

\subsubsection{Contribution from clusters to the bright galaxies budget}

A rough estimate of the contribution of the bright cluster galaxies
($<$ M$^{*}$ +2) to the total bright ($<$ M$^{*}$ +2) galaxy budget
can be obtained, by combining the cluster galaxies LF (and its
measured evolution) with the known space density of clusters out to
$z\sim\! 1$.

Our results show that at least at magnitudes brighter than M$^{*}$ +2,
the cluster galaxies LF appears to evolve mainly by passive evolution
up to $z\simeq 1$. In addition, the normalization of the K band
galaxy LF of X--ray luminous clusters was found to be consistent with
the local one at least out to $z \simeq 0.8$
\citep{trenthamemobasher1998}.  Therefore, we can assume that
high--redshift clusters of a given X--ray luminosity contain similar
numbers of bright ($<$ M$^{*}$ +2) galaxies as low--redshift clusters
of the same X--ray luminosity.

We use the relation between the number of galaxies brighter than
M$^{*}$ +2 within $r_{200}$, and the cluster mass within $r_{200}$, as
derived by Popesso et al. (in preparation):
\begin{equation}
N_{galaxies}\simeq 10^{-7.12}(M_{200}/M_{\odot})^{0.61}
\end{equation}

and the relation between the X--ray luminosity and the cluster mass
within $r_{200}$ \citep{reiprich2002}, in the form:
\begin{equation}
log\left[ \frac{L_{X(0.1-2.4 keV)}}{h_{50}^{-2} 10^{40} \mbox{erg s}^{-1}   } \right] \simeq -19 +1.58 log \left(\frac{M_{200}}{h_{50}^{-1} M_{\odot}}\right).
\end{equation}

By combining these two scaling relations (and converting to H$_{0}$ =
70 km s$^{-1}$ Mpc$^{-1}$, and $L_{X}(0.5-2.0 keV) [10^{44} \mbox{erg
s}^{-1}]$), one obtains the number of bright galaxies in a cluster of
luminosity $L_{X}$:

\begin{equation}
N(L_{X}) \simeq 74.5 \cdot (L_{X}(0.5 - 2.0 keV)[10^{44} \mbox{erg s}^{-1}])^{0.386}.
\end{equation}

In the assumption that this holds in the redshift range [$0\div 1$],
we can use the redshift evolution of the cluster X--ray luminosity
function \citep{rosati2002,mullis2004} to estimate the number density
of clusters at a given luminosity and a given redshift. By using the
N($L_{X}$) relation above, we can compute the number density of bright
galaxies in the cluster virial regions at each redshift. A comparison
with the field LF (we used \citet{trentham2005} at $z\simeq0$, and
\citet{dahlen2005} at $z\simeq 1$) yields the contribution of
cluster galaxies to the total bright galaxy budget.  We thus find that
a fraction of less than 10\% ($\simeq 6\div7 \%$) of the bright
galaxies at $z\simeq 0$ is located in the virial regions of X--ray
luminous clusters (in broad agreement with what reported in
\citet{depropris2003}), and similar ($\simeq$ 5\%) at $z \simeq 1$.

\section{Mass to light ratios}

The cluster dynamical mass--to--light (M/L) ratio has long been
a matter of interest due to the fact that clusters collapse from
regions several Mpcs wide, thus carrying both their mass and galaxy
content from representative portions of the Universe. Their M/L ratio
could thus be similar to that of the whole Universe, even thought this
is much dependent on how different galaxy evolution is in such high
density environments with respect to average density regions.\\

The cluster M/L ratio can in principle be used to study how galaxies
evolve in dense environments. The M/L ratio is known to increase with
the system mass going from galaxies to clusters \citep{bahcall1995},
in agreement with the predictions of models of biased galaxy formation
\citep{davis1985,bardeen1986}.  However, it is not yet completely
clear, also in recent studies, whether a mass dependence of the M/L
ratio is observed in the limited mass range of groups and clusters.
While some works find a measurable increase of the M/L ratio with the
total mass of the system (for instance \citet{schaeffer1993,
adami1998a,adami1998b, girardi2000,
hoekstra2001,girardi2002,marinoni2002,bahcall2002,lin2003,lin2004,rines2004,tully2005}),
some others conclude that the M/L ratio is approximately the same in
groups and clusters (for instance \citet{dressler1978, david1995,
carlberg1996,carlberg1997,cirimele1997, hradecky2000}).

As observed for instance in \citet{tully2005}, galaxy groups with
lower B band M/L ratios compared to more massive systems generally
have a population of late type galaxies with on--going star
formation. Moreover, systems with predominant E/S0/Sa population tend
to have larger M/L ratios; this would be consistent with more dense
regions forming earlier, and therefore ending their star formation
earlier.  More recently, the K band M/L ratio, having very little
dependence on on--going star formation, has been found to be
mass--dependent \citep{lin2003,lin2004}, with K band light per unit
mass being higher by a factor $\simeq 2$ in low mass clusters than in
massive ones.

As discussed for instance in \citet{rines2004}, the observed mass
dependence could be due either to processes like tidal stripping and
dynamical friction disrupting galaxies in massive clusters, or to a
reduced star formation efficiency in such systems (possibly the heating
of the ICM cutting off the supply of cold material needed to form
stars \citep{blanton1999, balogh2000}).

Keeping in mind these issues that complicate the comparison of
inhomogeneous samples, as well as possible systematic differences in
the masses estimated by different means (see for instance
\citet{sanderson2003,andernach2004}), we can compare the M/L ratio of
the clusters studied in this work with estimates at lower redshift.  A
comparison of the cluster M/L ratios in the B band, out to redshift
$z \simeq 0.8$, was presented for instance by \citet{hoekstra2002},
who found that the evolution of the M/L ratio is consistent with the
luminosity evolution of galaxies as derived from the fundamental plane
in distant clusters (see their figure 14).

\begin{table*} \centering  \caption{Luminosity function
parameters and estimated absolute K band luminosities and
mass--to--light ratios. Column 2: indicative radius within which the
LF and the M/L ratio are measured; column 3: original passband in
which the LF is measured; column 4: LF characteristic magnitude
M$^{*}$ as measured in the original passband, column 5: M$^{*}$
k--corrected to the restframe K$_{s}$ band (F160W was previously
corrected to K$_{s}$ band by a factor -0.5); column 6: LF faint end
slope $\alpha$ (note that for Cl0910 $\alpha$ is unconstrained, and
the error on M$^{*}$ for this cluster is estimated assuming that
$-1.4< \alpha <-0.4 $); column 7: the total restframe K$_{s}$
luminosity within the effective area listed in column 6 of table
\ref{tab:data}; column 8: the restframe K$_{s}$ band mass--to--light
ratio (the errors come from the errors on the projected mass and on
the total luminosity).  Note that for Cl1252 we report the total
luminosity and the M/L ratio derived from both the LFs from this work
and from T04 (first and second value respectively).
\label{tab:LFs} }
    \begin{tabular}{@{}l l l l l l l l@{}}
\hline
Cluster&r&passband&M$^{*}_{obs}$&K$^{*}_{rest}$&$\alpha$&$L_{Ks,<r}$&$(M/L)_{Ks,<r}$\vspace{0.06cm}\\
 &kpc & & AB mag&AB mag& &$10^{12} L_{\odot}$&$M_{\odot} / L_{\odot}$\vspace{0.06cm}\\
\hline
RDCS J0910+5422&600&K$_{s}$&$20.4^{+1.2}_{-1.1}$&$ -23.27$&$-0.85$&$10^{+8}_{-1}$&$40^{+20}_{-30}$ \vspace{0.06cm}\\
RDCS J1252.9-2927&500&K$_{s}$&$20.1^{+1.1}_{-1.2}$&$-23.89$&$-0.95^{+0.35}_{-0.35}$&$ 18^{+5}_{-6} \div 14^{+3}_{-3}$&$ 13^{+5}_{-5} \div 16^{+5}_{-5}$   \vspace{0.06cm}\\
RX J0848+4453&400&F160W ($\simeq$ H)&$20.8^{+1}_{-1.6}$&$ -23.76^{+1}_{-1.6}$&$-1.15^{+0.4}_{-0.3}$ &$ 4^{+2}_{-1}$&$35^{+25}_{-30}$ \vspace{0.06cm}\\
composite LF & $-$ & K$_{s}$ &$20.5^{+0.4}_{-1}$  & $-23.41^{+0.4}_{-1}$&$-1.0^{+0.2}_{-0.3}$& $-$ & $-$\vspace{0.06cm}\\
\hline
     \end{tabular}
\end{table*}

We have estimated the K band M/L ratios of Cl0910, Cl1252, Cl0848 making use
of the LFs derived above and of the X--ray mass profiles derived in
\citet{ettori2004}.

Once the Schechter parameters $\alpha$ and K$^{*}_{rest}$ (and the
corresponding characteristic luminosity $L^{*}$) have been determined,
the K band projected total luminosity within the surveyed area can be
calculated via direct integration of the Schechter function
($L_{tot}=\Phi^{*} L^{*} \Gamma (2+\alpha)$). The K band luminosities
for the three clusters within the surveyed area (as listed in column 6
of table \ref{tab:data}) are listed in table \ref{tab:LFs}.  We assume
K$_{\odot}$ = 5.2 when calculating luminosities in units of
$L_{\odot}$.  The errors were determined by calculating the luminosity
with Schechter parameters running on the 1--$\sigma$ confidence level
for K$^{*}$ and $\alpha$.

The projected M/L ratios for the three clusters were derived within
the surveyed area, using the projected mass profiles derived in
\citet{ettori2004} (see table \ref{tab:LFs}).

Note that the mass to light ratio derived for Cl1252 by summing up the
luminosities of the photometric members from T04 without further
corrections ($M/L_{K} = 17^{+3}_{-3}M_{\odot}/L_{\odot}$) is
consistent with the quoted value. By considering only the confirmed
spectroscopic members, we obtain $M/L_{K} = 31\pm5 M_{\odot}/L_{\odot}$,
which should be considered as an upper limit.

It should be noted that, since our surveyed areas are different from
each other and are small, we rely on the assumption of a negligible
or small dependence of the M/L ratio on the clustercentric distance within the virial radius
\citep{rines2001,kneib2003,rines2004}, when comparing the M/L ratios
derived here with other measurements.

\begin{figure}
\centering
\includegraphics[width=8.8cm]{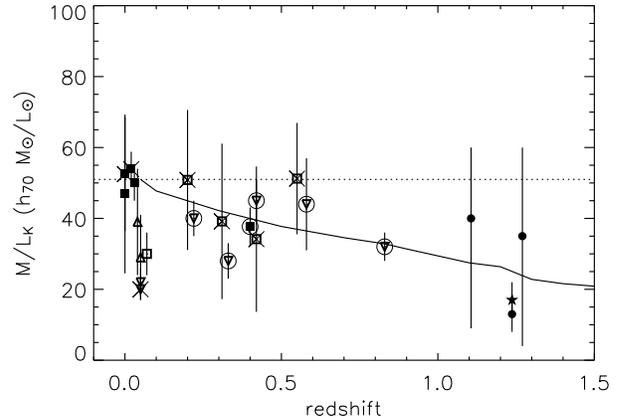}

\caption{The evolution of the restframe K band mass--to--light
ratio. The filled circles are the M/L ratios from this work. The star
shows the M/L determined for Cl1252 using the T04 LF. The filled
squares are K band M/L measurement from
\citet{rines2001,kneib2003,lin2003,rines2004}. The empty
symbols show the K band M/L ratio derived from published M/L estimates
in other passbands, i.e. B/B$_{j}$ band (circle) from
\citet{girardi2000,hoekstra2002,sanderson2003,gavazzi2004}, R/r band
(square) from \citet{carlberg1997} and \citet{andernach2004}, and
V band from \citet{cirimele1997} and \citet{hradecky2000} (see
text). Crossed and circled points have masses determined from galaxy
kinematics or strong/weak lensing respectively, at a difference with
X--ray masses used in this work. The solid line traces the evolution
expected from pure luminosity evolution of a simple stellar population formed at z=5, assuming that the cluster M/L ratio
only evolves because of stellar evolution in galaxies, and neglecting any
evolution in the dark mass. The dotted line shows the reference (z=0) value.
\label{fig:MLK}}
\end{figure}

In figure \ref{fig:MLK} the K band M/L ratio for the three clusters is
compared to previous determinations at lower redshifts
\citep{carlberg1997,cirimele1997,girardi2000,hradecky2000,rines2001,hoekstra2002,kneib2003,lin2003,sanderson2003,andernach2004,gavazzi2004,rines2004}.

The M/L ratios published in passbands different from $K$ were rescaled
to the $K$ band using the colours of a simple stellar population formed at z=5
and the AB colours of the Sun $(B-K)_{\odot}$=0.15, $(R-K)_{\odot}$ =
-0.76, $(V-K)_{\odot}$ = -0.33. These rescaled measurements are shown
with empty symbols.

To avoid excessive confusion in the plot, some points do not represent
a single cluster but are based on different samples: 32 groups and
clusters from \citet{sanderson2003}, 8 groups and clusters from
\citet{hradecky2000}, 16 clusters from \citet{carlberg1997} grouped in
4 redshift bins, 105 clusters from \citet{girardi2000}, 12 clusters
from \citet{cirimele1997}, 13 clusters from \citet{lin2003}, 180
clusters from \citet{andernach2004}, and 9 clusters from
\cite{rines2004}. In such cases, the weighted average (and
corresponding error) of the sample is plotted in figure \ref{fig:MLK}.

Some of the M/L ratios plotted in figure \ref{fig:MLK} were derived
with masses estimated from kinematics (crossed points) or from
strong/weak lensing (circled points).  The solid line traces the
expected evolution of the M/L ratio, neglecting any evolution in the
dark halo mass, in the assumption that the M/L ratio evolves following
the luminosity evolution of the cluster galaxies, for a pure
luminosity evolution of a simple stellar population formed at $z=5$
(normalized at M/L = 51 at redshift zero).

\section{The stellar mass function}

At redshift $\simeq 1$, the K$_{s}$ band (rest-frame 1$\mu$m)
luminosity is still a good tracer of the stellar mass. Therefore, we
can draw an estimate of the stellar mass function (MF) of cluster
galaxies at redshift $z \simeq 1.2$ from the composite K$_{s}$ band
LF.  The K$_{s}$ band light translates into the stellar mass via the
stellar M/L ratio of each galaxy's stellar population, which depends
on the galaxy star formation history (SFH) and on its age.

In principle, if we knew which galaxies are contributing to the
composite LF (and we measured their photometry in a sufficient number
of passbands) we could measure the stellar M/L ratio for each of the
galaxies via SED fitting, and directly determine the stellar mass
function.  However, since the composite LF was determined in a
statistical fashion, we also have to statistically evaluate the
stellar M/L ratios of cluster galaxies along the LF.  To this aim, we
used the Cl1252 photometric members selected by T04 to statistically
estimate, in each magnitude bin of the LF, the median and scatter
(16$^{th}$--84$^{th}$ percentiles) colours i$_{775}$-K$_{s}$,
z$_{850}$-K$_{s}$, and J$_{s}$-K$_{s}$.  In the assumption that the
contributions to the LF from different galaxy populations in Cl0910
and Cl0848 are approximately similar to those in Cl1252, these colours
obtained along the Cl1252 LF will be representative of those along the
composite LF.

We then built a set of 160 synthetic SEDs with the
\citet{bruzualecharlot} code, with (delayed) exponentially declining
SFH with $0.005 <\tau< 1.5$ Gyr and $0.3 <$ age $< 5$ Gyr, solar
metallicity, Salpeter IMF and no reddening. For each magnitude bin, we
selected all the models with the appropriate colour (median $\pm$
scatter as above) for that bin, thus determining a rough estimate of
the M/L ratios in that magnitude interval.
For each of the three colours i$_{775}$-K$_{s}$,
z$_{850}$-K$_{s}$, and J$_{s}$-K$_{s}$ we extracted 100 realizations
of the MF perturbing the number of galaxies in each magnitude bin with
a $\sigma$ equal to its Poissonian error and then spreading in mass
this perturbed number of galaxies within the above defined M/L ratio
range for that magnitude bin. We then considered the median and
minimum/maximum MFs over the 100 realizations, for each of the three
colours, which give three estimates of the MF which are perfectly consistent.

\begin{figure}
\centering
\includegraphics[width=8.8cm]{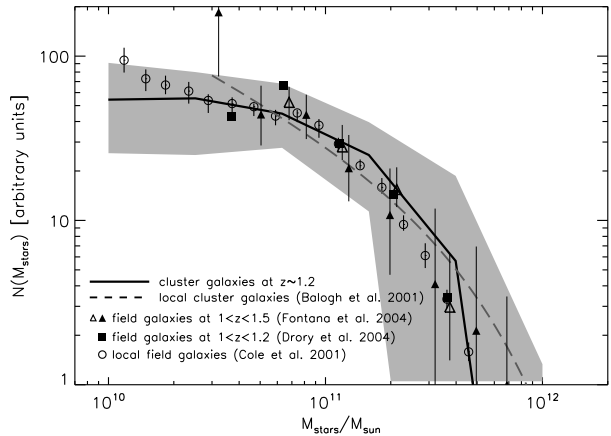}
\caption{The stellar mass function as determined from the composite K$_{s}$
  band LF. The shaded area accounts for errors in the K$_{s}$ band LF
  and in the stellar mass to K$_{s}$ light ratio (see text for
  details). Other determinations of the stellar mass function are
  shown for reference, as indicated in the legend.
\label{fig:massF}}
\end{figure}

We thus averaged these three estimates obtaining the MF plotted in
figure \ref{fig:massF}; the shaded region corresponds to the average
minimum/maximum MFs computed according to the above description.  We
also show for comparison the MFs for field galaxies at similar
redshift \citep{drory2004,fontana2004}, for field galaxies at redshift zero
\citep{cole2001}, and for local cluster galaxies
\citep{balogh2001}\footnote{The MF from \citet{balogh2001} was
calculated with a Kennicutt IMF, so we applied a correction logM$_{\rm
stars}^{salpeter}$ = logM$_{\rm stars}^{kennicutt}$ +0.35.}, all
arbitrarily rescaled so to match at M$_{\rm stars} \simeq 5 \cdot
10^{10} M_{\odot}$. The MF is shown for masses greater than $ \simeq
10^{10} M_{\odot}$, which is the estimated mass completeness
determined for our completeness magnitude, considering the M/L ratio
of a stellar population formed at z$\simeq$10 with subsolar
metallicity and no reddening.  The stellar mass corresponding to the
characteristic magnitude K$_{s}^{*} = 20.5$ is approximately  $10^{11}$ M$_{\odot}$.

Within the uncertainties affecting this MF determination, the shape of
the MF of massive objects at redshift $\simeq 1.2$ is not
significantly different from the local one as measured from the 2dF
galaxies \citep{cole2001}. This is in agreement with the very mild
evolution of the mass function observed in the redshift range [0:1]
both in the field (e.g., \citet{fontana2004}) and in clusters (e.g.,
\citet{kodamaebower}).  The presence of massive objects (M$_{\rm stars} \geq
10^{11} M_{\odot}$) is independently confirmed at least in Cl1252 from
SED fitting on 9 passbands from restframe NUV to NIR (Rosati et al., Rettura et al.,
in preparation). However, we note that by probing galaxy clusters we
are only sensitive to the evolution of the shape of the MF, and not to
its normalization, whereas field galaxy surveys find the evolution of
the MF to become significant at $z \geq 1$
\citep{fontana2004,drory2004}.

\section{Conclusions}

We have studied the near infrared luminosity function of high redshift
cluster galaxies in three X--ray luminous clusters that are among the
most distant discovered so far ($1.1<z<1.3$).

These clusters bear the strongest leverage on evolutionary studies as
they probe a redshift regime when the Universe was less than half of
its present age.  By measuring the K band luminosity function
(and galaxy stellar mass function) in these systems, and by comparing
it with that at $z \simeq 0$, we can set valuable constraints on the
galaxy evolution in dense environments in the redshift range [0 $\div$
1].

We should note, however, that derived quantities still suffer from
large uncertainties, due to small--number statistics and possibly
biases in the galaxy populations (due to probing only the cluster
cores).

The LF resulting from this work is consistent with previous
determinations at similar or lower redshifts, and consolidates a
scenario where the evolution of the characteristic magnitude M$^{*}$
($\Delta M^{*} \simeq -1.3$) is consistent with predictions of passive
evolution for a stellar population formed at $z >2$. Moreover, we find
that the overall shape of the high redshift LF matches the one of the
local cluster galaxies LF, once such a brightening is taken into
account.

Similarly, the evolution of the K band LF of field galaxies has
been found to be consistent with passive evolution up to $z = 1$, with
a density evolution lower than 30\% and a brightening of $\Delta M^{*}
\simeq 0.5 \div 0.7$ mag (e.g., \citet{pozzetti2003,drory2003} --
however, see also \citet{dahlen2005}).

A direct comparison of the field and cluster LFs at redshift zero, has
revealed a significant difference both in the B and in the K band
(e.g. \citet{balogh2001,depropris2003}) amounting to $\simeq 0.3$ mag
in M$^{*}$ and $\simeq 0.1\div 0.2$ in $\alpha$. While there is a hint
of exceeding very bright galaxies in our distant clusters with respect
to the field, our error bars do not allow us to make such a claim.

For Cl1252, for which best quality data and extended wavelength
coverage are available, we attempted a separation of the contributions
of late and early type galaxies to the bright end of the
LF. We find that, already at $z \simeq 1.2$, the bright end of the LF
appears to be  dominated by early type galaxies selected either on the basis
of their morphological appearence or of their spectrophotometric
properties.

Using the individual LFs for the three clusters, we calculated the
K band cluster M/L ratios making use of X--ray mass profiles derived
in \citet{ettori2004}. The M/L ratio tends to be smaller than the
typical value at redshift zero, as expected on the basis of pure
luminosity evolution of the cluster galaxies stellar populations.
However, a much larger sample would be needed for a detailed
investigation of the evolution of the cluster M/L ratio.

Finally, from the composite K$_{s}$ band LF we have estimated the
stellar mass function of cluster galaxies.  The observed K$_{s}$ band
light at $z \simeq 1.2$ corresponds approximately to the restframe
{\it z} band light, and is considered as a good tracer of the stellar
mass.  We have outlined in the introduction how the determination of
the stellar mass function at high redshift sets strong constraints on
the evolution of massive galaxies.  While the early formation epoch of
the bulk of the stars in massive galaxies is now generally established
by several observations, the epoch of the major mass assembly can only
be assessed by studying the redshift evolution of the mass function.
Our study shows that the massive ($M_{stars} > 10^{10} M_{\odot}$)
galaxy populations in massive high redshift clusters have not
significantly changed since $z \simeq 1$, apart from passive evolution
of their stars, thus extending previous results at lower redshifts.
The shape of the stellar mass function at $z \simeq 1.2$ is found to
be consistent with the one observed in local clusters, within our
uncertainties. 

The high-mass end of the LF, made of giant ($M_{\rm stars}>10^{11} M_{\odot}$)
E/S0 galaxies, is already in place at $z\simeq 1.2$.  This points
toward an early assembly of the galaxy mass, mostly completed before
$z \simeq 1$, thus implying that the bulk of merging activity for
massive galaxies in clusters has to occur at much earlier epochs.

This might appear not surprising as field studies have found evidence
of a similar early assembly of massive galaxies with most of the
stellar mass already assembled in systems more massive than the local
characteristic mass by redshift 1 (e.g., \citet{fontana2004,conselice2005}).
Since galaxy evolution in clusters is expected to be faster than in
the field in hierarchical galaxy formation scenarios, an even milder
evolution of the mass function of cluster galaxies is expected.  The
similarity between  the shape of the MF we have found at $z\simeq 1.2$ in
rich clusters and in the field sets upper limits on the difference of
formation time--scales of stellar populations in high and low density
environments.

For example, if we use the star formation histories for
early type galaxies in different environments, for different stellar
masses, as derived from the fossil record data \citep{thomas2005}, our
results would imply an age difference of $\leq 2$Gyr.

The situation is probably different for
lower mass galaxies, however it remains difficult to probe the
mass function significantly lower than $10^{10} M_{\odot}$ at $z>1$.

Among the caveats in our study, we should mention the so--called
progenitor bias. Since we have considered the whole cluster galaxy
population, our work is in principle not affected by the progenitor
bias referred to when dealing with selected populations of galaxies
(generally early--types \citep{vandokkum2001b}). It remains true that
galaxy populations in clusters at high redshift might not be directly
comparable to local ones (e.g., \citet{kauffmannecharlot1998a}), and
that the high--redshift clusters we are observing might not be the
progenitors of the local X--ray luminous clusters
\citep{kauffmann1995}, leading to an underestimated evolution.

With the aid of multicolour photometry, including Spitzer/IRAC bands,
we can now directly estimate the stellar masses of high redshift
cluster galaxies, as well as approximate ages of their stellar
populations, and push these studies out to $z=1.4$
(\citet{mullis2005}, Stanford et al. (2005)), thus
probing an epoch which is thought to be crucial for the formation of
massive clusters. This work will stimulate significant progress in
discriminating between different formation scenarios.

\begin{acknowledgements}
  We thank T. Kodama for providing us with his elliptical galaxy
  evolution models, and P. Popesso for sharing her results prior to
  publication. We also thank the anonymous referee for useful comments
  which improved the presentation of this work.  VS thanks G. De
  Lucia, M. Esposito, M. Pannella, M.  Paolillo, and A. Rettura for
  helpful discussions and comments.  VS gratefully acknowledges
  support from the European Social Fund through a PhD grant, and from
  the ESO Director General Discretionary Fund program.  ST received
  support from the Danish Natural Science Research Council.  This work
  made use of observations of the GOODS--S field carried out using the
  Very Large Telescope at the ESO Paranal Observatory under Program
  ID: LP168.A-0485.
\end{acknowledgements}

\bibliography{KLFsub3_astroph}

\end{document}